\documentclass[%
 reprint,
superscriptaddress,
 amsmath,amssymb,
 aps,
]{revtex4-2}

\usepackage{graphicx}% Include figure files
\usepackage{dcolumn}% Align table columns on decimal point
\usepackage{bm}% bold math
\usepackage{xspace}
\usepackage{graphicx}% Include figure files
\usepackage{dcolumn}% Align table columns on decimal point
\usepackage{bm}% bold math
\usepackage{xspace}
\usepackage{natbib} % APS 样式强依赖 natbib

\usepackage{hyperref}% 必须启用它，因为 \Eprint 强烈依赖 hyperref 的超链接功能

\begin{document}

\preprint{APS/123-QED}

\title{Little red dots as a cosmological probe: constraining $H_0$ with quasi-periodic pulsations}% Force line breaks with \\
\newcommand{\mbh}{M_{\rm BH}}
\newcommand{\msun}{M_\odot}
\newcommand{\rsun}{R_\odot}
\newcommand{\zsun}{Z_\odot}
\newcommand{\lsun}{L_\odot}
\newcommand{\gcc}{{\rm g~cm}^{-3}}
\newcommand{\cc}{{\rm cm}^{-3}}
\newcommand{\mdot}{\dot m}
\newcommand{\msunyr}{M_\odot~{\rm yr}^{-1}}
\newcommand{\mdote}{\dot{\rm M}_{\rm Edd}}
\newcommand{\kpc}{{\rm kpc}}
\newcommand{\mpc}{{\rm Mpc}}
\newcommand{\gpc}{{\rm Gpc}}
\newcommand{\pc}{{\rm pc}}
\newcommand{\mum}{\mu {\rm m}}
\newcommand{\kms}{{\rm km~s}^{-1}}
\newcommand{\ergs}{{\rm erg~s}^{-1}}
\newcommand{\ledd}{{\rm L}_{\rm Edd}}

\newcommand{\Muv}{M_{\rm UV}}
\newcommand{\SFR}{{\rm SFR}}
\newcommand{\yr}{\rm yr}
\newcommand{\Hmol}{\rm H_2}
\newcommand{\K}{{\rm K}}
\newcommand{\beq}{\begin{equation}}
\newcommand{\eeq}{\end{equation}}
\newcommand{\muJy}{\mu{\rm Jy}}
\newcommand{\D}{{\rm d}}
\newcommand{\Om}{\Omega_{\rm m}}

\newcommand{\red}[1]{\textcolor{red}{ #1}}
\newcommand{\blue}[1]{\textcolor{blue}{ #1}}
\newcommand{\cyan}[1]{\textcolor{cyan}{ #1}}
\newcommand{\rxc}{RXC\,J2211--0350\xspace}
\newcommand{\tgta}{R2211-RX1\xspace}
\newcommand{\tgtb}{R2211-RX2\xspace}
\newcommand{\ion}[2]{#1\,\textsc{\lowercase{#2}}}

\author{Zijian Zhang}
% \email{ } 
\affiliation{%
 Kavli Institute for Astronomy and Astrophysics, Peking University, Beijing 100871, People's Republic of China
}%
\affiliation{%
 Department of Astronomy, School of Physics, Peking University, Beijing 100871, People's Republic of China
}%

\author{Kohei Inayoshi}
\email{inayoshi@pku.edu.cn}
\affiliation{%
 Kavli Institute for Astronomy and Astrophysics, Peking University, Beijing 100871, People's Republic of China
}
\affiliation{%
 Department of Astronomy, School of Physics, Peking University, Beijing 100871, People's Republic of China
}%

\author{Masamune Oguri}
\affiliation{%
 Center for Frontier Science, Chiba University, 1-33 Yayoi-cho, Inage-ku, Chiba 263-8522, Japan
}%
\affiliation{%
 Department of Physics, Graduate School of Science, Chiba University, 1-33 Yayoi-Cho, Inage-Ku, Chiba 263-8522, Japan
}%

\author{Linhua Jiang}
\email{jiangKIAA@pku.edu.cn}
\affiliation{%
 Kavli Institute for Astronomy and Astrophysics, Peking University, Beijing 100871, People's Republic of China
}%
\affiliation{%
 Department of Astronomy, School of Physics, Peking University, Beijing 100871, People's Republic of China
}%

\author{Fengwu Sun}
% \email{fengwu.sun@cfa.harvard.edu}
\affiliation{%
 Center for Astrophysics $|$ Harvard \& Smithsonian, 60 Garden St., Cambridge, MA 02138, USA
}%

\author{Mingyu Li}
% \email{lmy22@mails.tsinghua.edu.cn}
\affiliation{%
 Department of Astronomy, Tsinghua University, Beijing 100084, People's Republic of China
}%

\author{Xiaojing Lin}
\affiliation{%
 Department of Astronomy, Tsinghua University, Beijing 100084, People's Republic of China
}

\date{\today}% It is always \today, today,

\begin{abstract}
The James Webb Space Telescope (JWST) has uncovered a population of ``little red dots'' (LRDs) at $z \gtrsim 4$, potentially representing early supermassive black holes embedded in dense gaseous envelopes.
The recent discovery of the lensed LRD \tgta\ reveals significant variability on rest-frame timescales of decades, which may be interpreted as quasi-periodic variation that has a potential physical parallel to stellar pulsations.
In this work, we derive an idealized, self-consistent period-luminosity-temperature ($P$-$L$-$T_{\rm eff}$) relation based on the hydrostatic envelope model.
If this theoretical relation holds and can be empirically validated/calibrated, it would offer a novel framework for constraining the Hubble constant ($H_0$).
The current sparse sampling of \tgta\ yields a preliminary $H_0 = 120.7_{-46.5}^{+47.0} \text{ km s}^{-1}\text{ Mpc}^{-1}$ as a proof-of-concept, with the error budget dominated by the uncertainty of the pulsation period.
Our forecasting analysis shows that continuous monitoring over a 10-year baseline can reduce the $H_0$ uncertainty to 3–20\%, depending on the intrinsic pulsation period, while the systematic uncertainty floor remains to be fully characterized.
This method offers a potential independent probe to measure luminosity distances in the early universe.
\end{abstract}
% \begin{abstract}
% ``Little red dots'' (LRDs) at $z \gtrsim 4$ discovered by JWST may represent early supermassive black holes embedded in dense gaseous envelopes.
% Recent observations of the lensed LRD \tgta\ reveals significant decade-scale variability that can be explained as a quasi-periodic variation analog to stellar pulsations.
% We derive an self-consistent period-luminosity-temperature relation for hydrostatic envelope and provide a novel framework to constrain the Hubble constant $H_0$.
% A proof-of-concept analysis for \tgta\ yields $H_0 = 120.7_{-46.5}^{+47.0}~\rm km~s^{-1}~Mpc^{-1}$, with the error currently dominated by the period measurement uncertainty.
% With robust control of systematic uncertainties, a ten-year monitoring will achieve a 3–20\% precision in $H_0$, enabling LRDs as an independent cosmological probe.
% \end{abstract}

\maketitle

%\tableofcontents

\section{\label{sec:}Introduction}

James Webb Space Telescope (JWST) has revealed a ubiquitous population of compact \citep{Furtak_2023a,Akins_2025}, high-redshift ($z \gtrsim 4$) sources characterized by unique V-shaped spectral energy distributions (SEDs), widely known as ``little red dots'' \citep[LRDs;][]{Barro_2024,Matthee_2024,Labbe_2025}. 
While their compactness and broad Balmer emission lines suggest accreting supermassive black holes \citep[SMBH;][]{Harikane_2023_agn,Greene_2024,Matthee_2024,Wang_2024b,Kocevski_2025}, LRDs also exhibit puzzling characteristics compared with typical active galactic nuclei (AGNs), such as the unusual faintness in X-ray \citep[][]{Ananna_2024,Yue_2024,Maiolino_2025}, radio \citep[][]{Mazzolari_2024}, and observed mid-to-far infrared bands \citep[][]{Perez-Gonzalez_2024,Williams_2024,Akins_2025}, and the little to no short-term variability \citep[][]{Kokubo_Harikane_2024,Tee_2025,Zhang_2025,Liu_2026_twinkle}. 
They may represent a transitional or previously unexplored phase of BH growth in the early universe \citep[][]{Inayoshi_2025a}.

Compelling evidence increasingly suggests that LRDs are AGNs cocooned within dense, optically thick gaseous envelopes \citep{Inayoshi_Maiolino_2025,Ji_2025,Kido_2025,Naidu_2025}. 
In this scenario, the central AGN radiation is reprocessed with an envelope with a surface temperature of $T_{\rm eff} \simeq 4000$--$6000$ K, similar to the typical photospheric temperature of giant stars. This dense gas envelope naturally explains the red optical continua, prominent Balmer breaks, and a high fraction of absorption lines on LRD spectra \citep[][]{Lin_2024,Wang_2024b,deGraaff_2025a,Juodzbalis_2024,Naidu_2025}. 
This ``BH-envelope'' hypothesis has gained significant support from the discovery of local LRD analogs at $z \simeq 0.1$--$0.2$, like J1025$+$1402 \citep[][]{Ji_2026,Lin_2026}. It displays the full set of LRD characteristics, together with strong absorption features from low-ionization metal species such as the \ion{Ca}{II} triplet, \ion{Na}{I} D, and \ion{K}{I}. 
These spectral fingerprints suggest that the gas properties at the LRD envelope surface are remarkably similar to those of yellow supergiants.

The BH-envelope model naturally predicts quasi-periodic variability driven by $\kappa$-mechanism pulsations in radiation-pressure-dominated\citep{Cantiello_2025,Inayoshi_2026b}, physically analogous to stellar pulsations giant stars \citep[][]{Eddington_1917,Cox_1980,Li_1994} and in accreting massive stars in the present-day universe \citep{Inayoshi_2013b,Pandey_2025}, a short-lived stage before collapsing into heavy BH seeds \citep{Hosokawa_2013,Inayoshi_2013}. 
% A natural prediction of the BH-envelope model is the existence of quasi-periodic variability \citep{Cantiello_2025,Inayoshi_2026b}. 
% Much like stellar pulsations driven by the $\kappa$-mechanism in giant stars such as Cepheids \citep[][]{Eddington_1917,Cox_1980,Li_1994} and in accreting massive stars in the present-day universe \citep{Inayoshi_2013b,Pandey_2025}, the radiation-pressure-dominated gaseous envelopes of LRDs are expected to undergo periodic pulsations in analogy of accreting supermassive stars, a short-lived stage before collapsing into heavy BH seeds \citep{Hosokawa_2013,Inayoshi_2013}. 
% \citet{Zhang_2025c} recently provided evidence for such variability through a century-long baseline study of a gravitationally lensed LRD with quadruply multiple images, \tgta\ in the galaxy cluster \rxc. By leveraging the $\sim 130$-year time delay between its lensed images, they detected intrinsic brightness and color variations consistent with blackbody temperature and radius fluctuations of a large photosphere ($R \sim 2000$ AU) with a rest-frame period of decades. 
\citet{Zhang_2025c} recently provided evidence for such variability in a gravitationally lensed LRD \tgta, where $\sim 130$-year time delay between images reveal intrinsic brightness and color variations consistent with blackbody temperature and radius fluctuations over a rest-frame period of decades. 
The pulsation nature of this variability is further supported by the instability analysis of \citet{Cantiello_2025}, who demonstrated that such gas envelopes surrounding an accreting BH \citep[i.e., a quasi-stellar structure;][]{Begelman_2006} are unstable to radial pulsations driven by the $\kappa$-mechanism within a specific instability strip.
% This pulsational nature is supported by instability analysis of quasi-stellar envelopes \citep[e.g., ][]{Begelman_2006}, which undergo $\kappa$-mechanism-driven radial pulsation \citep{Cantiello_2025}.
% This discovery highlights the physical parallels between the variability of some LRDs and stellar pulsations of giant stars.
% The existence of pulsations in some LRDs may open a new window to constrain cosmology at high redshifts. 
% Modern cosmology is built upon the ($\Lambda$ cold dark matter) $\Lambda$CDM framework, where the Hubble constant ($H_0$) is a fundamental parameter representing the present expansion rate of the Universe. However, a significant tension has emerged between the $H_0$ values inferred from early-universe probes, such as the cosmic microwave background \citep[CMB;][]{Planck_2020}, and late-universe measurements like Type Ia supernovae \citep{Riess_2022_SHOES}. This ``Hubble tension'' signals potential unresolved systematic effects or new physics beyond the standard model \citep{DiValentino_2021}, highlighting the urgent need for independent cosmological probes.

Such pulsations may offer a novel window to constrain the Hubble constant ($H_0$), helping to address the emerging ``Hubble tension'' between early- and late-universe measurements \citep{Planck_2020,Riess_2022_SHOES,DiValentino_2021}.
This tension signals potential unresolved systematic effects or new physics beyond the standard model \citep{DiValentino_2021}, highlighting the urgent need for independent cosmological probes.
If pulsating LRDs follow a robust period-luminosity ($P$-$L$) relation analogous to the Leavitt Law for Cepheids \citep{Leavitt_1912}, they could serve as independent ``standard candles'' to measure luminosity distance at high redshifts.
% Unlike traditional empirically calibrated standard candles, this approach is initially grounded in a physical analogy to pulsating stars within the BH–envelope framework.
In this Letter, we establish an idealized, self-consistent $P-L-T_{\rm eff}$ framework for LRD pulsations and derive its direct relation to $H_0$. We then apply this model to \tgta\ and demonstrate through forecasting that a modest sample of LRDs can constrain $H_0$ to within $\lesssim 10\%$, providing a unique cosmological probe at high redshifts.

\section{Physical Model and Cosmological Inference Framework}
\label{sec:lrd_pulsation}

\subsection{LRD Pulsation Model}

The purpose of this subsection is not to present a competitive measurement of the Hubble constant.
We instead aim to establish a self-consistent theoretical framework linking pulsation observables of LRDs to cosmological distances and to identify 
the dominant sources of uncertainty relevant for future observations.

Some LRDs exhibit deterministic variability consistent with radial pulsation, which can be described by
\begin{equation}
P_{\rm src} = 2\pi Q \left(\frac{R^3}{G M}\right)^{1/2},
\label{eq:pulsation}
\end{equation}
where $M$ and $R$ denote the BH mass and the photospheric radius.
$Q$ is the pulsation constant characterized by the mechanical and thermal structure of the envelope (see below).
Since the photospheric emission is powered by mass accretion onto the central BH, we parameterize the time-averaged luminosity of the radiation-pressure-supported envelope in quasi-hydrostatic equilibrium as
\begin{equation}
L = (1-\beta) L_{\rm Edd} = (1-\beta) \frac{4\pi cG M}{\kappa_{\rm es}},
\label{eq:eddington}
\end{equation}
where $\kappa_{\rm es}$ is the Thomson scattering opacity and $\beta=P_{\rm gas}/(P_{\rm gas}+P_{\rm rad})$.
For an envelope in hydrostatic equilibrium with a polytropic index $n=3$, one obtains 
$1-\beta \approx 0.003~(\mu \beta)^4 (M/\msun)^2$ \citep{Chandrasekhar_1939,Hoyle_1963}, leading to
\begin{equation}
\beta \approx 0.00256 \left(\frac{\mu}{0.6}\right)\left(\frac{M}{10^6~\msun}\right)^{-1/2}.
\end{equation}
Therefore, $\beta$ is not an independent, free parameter but is determined by the BH mass.
For such a radiation-pressure-dominated envelope, where the specific heat index approaches $\gamma_{\rm ad}\simeq 4/3+\beta/6$ and a constant entropy profile is maintained due to efficient convection, the pulsation constant is computed as
$Q=0.3330$ for the first overtone mode \footnote{
The pulsation period for the fundamental mode goes infinity because the restoring force in the limit of 
$\gamma_{\rm ad}\rightarrow 4/3$. However, non-zero contributions from gas pressure maintains $\gamma_{\rm ad}-4/3\simeq \beta/6>0$
and yields a finite period \citep{Inayoshi_2013}.} \citep{Schwarzschild_1941}.
\citet{Cantiello_2025} found that the pulsation constant is $Q=0.3226$
for the first overtone mode in the quasi-stellar configuration.
The two values are consistent at a $\sim 3\%$ error.
We adopt the latter value in the following analysis ($Q_0=0.3226$).

Assuming that the emission follows a blackbody spectrum with an effective temperature $T_{\rm eff}$, the luminosity is given by
\begin{equation}
L = 4\pi R^2 \sigma_{\rm SB} T_{\rm eff}^4,
\label{eq:sb}
\end{equation}
where $\sigma_{\rm SB}$ is the Stefan--Boltzmann constant.
% While the envelope undergoes radial pulsations where $R$ and $T_{\rm eff}$ fluctuate over time, Eqs. (\ref{eq:pulsation}) and (\ref{eq:eddington}) characterize the time-average $L$ and $T_{\rm eff}$ of the system. Eq. (\ref{eq:sb}) can describe the instantaneous state of the photosphere during the pulsation, and it also holds for the time-averaged values. 
While the envelope pulsates, Eqs. (\ref{eq:pulsation}) and (\ref{eq:eddington}) characterize its time-averaged state, with Eq. (\ref{eq:sb}) remaining valid for both instantaneous and average values.
Combining Eqs. (\ref{eq:pulsation})--(\ref{eq:sb}), we obtain a period--luminosity--temperature relation:
\begin{align}
P_{\rm src}
&=Q_0\left(\frac{4\pi^3 c^2}{\kappa^2\sigma_{\rm SB}^3}\right)^{1/4}(1-\beta)^{1/2}L^{1/4}T_{\rm eff}^{-3}, \label{eq:Psrc} \\
%&=30.2~{\rm yr}~Q_0
&=9.74~{\rm yr}~
(1-\beta)^{1/2}\left(\frac{L}{10^{10}~\lsun}\right)^{1/4}\left(\frac{T_{\rm eff}}{5000~\K}\right)^{-3}.
\nonumber
\end{align}
This decade pulsation period is accessible for relatively long-term monitoring,
especially for low-redshift LRDs (e.g., $z<1$).

Eq. (\ref{eq:Psrc}) provides an idealized theoretical baseline, though complexities—akin to classical pulsating variables—may arise. 
% Similar to classical Cepheid pulsating variables, the actual pulsation mechanism in LRD envelopes may exhibit complexities not fully captured by a simple polytropic model. 
Factors such as the pulsation constant $Q$ \citep{Cox_1980} and potential dependence on the envelope metallicity could introduce systematic deviations \citep[][]{Bond_1991,Macri_2006}. 
While future empirical calibration is essential to refine this relation into a precision cosmological tool, this model provides a robust approximation. In the following sections, we adopt this framework to demonstrate the potential of LRDs as an independent $H_0$ probe.

\subsection{Constraint on $H_0$}

\subsubsection{Direct Period Measurements}

The luminosity distance is related to the observed flux as $D_{L}=(L/4\pi F_{\rm obs})^{1/2}$, and thus can be expressed as
\begin{equation}
D_{L}=\frac{\kappa_{\rm es}F_{\rm obs}}{c(1-\beta)} 
\left(\frac{P_{\rm src}}{2\pi Q_0}\right)^{2} 
\left(\frac{\sigma_{\rm SB} T_{\rm eff}^4}{F_{\rm obs}}\right)^{3/2}.
\end{equation}
Using the definition of the luminosity distance and $P_{\rm obs}=(1+z)P_{\rm src}$, we obtain
\begin{equation}
H_0 = (1+z)^3\frac{c^2(1-\beta)}{\kappa_{\rm es}}
\cdot
\frac{\langle F_{\rm obs} \rangle^{1/2}}{\langle \sigma_{\rm SB} T_{\rm eff}^4 \rangle^{3/2}} 
\left(\frac{P_{\rm obs}}{2\pi Q_0}\right)^{-2}
\int_{0}^{z}\frac{dz'}{E(z')},
\label{eq:H0_direct}
\end{equation}
where all observed quantities are defined as time-averaged values over one pulsation period,
\begin{equation}
\langle A \rangle = \frac{1}{P_{\rm obs}}\int_0^{P_{\rm obs}}A(t) dt.
\end{equation}
This treatment is required because the relations above are defined with global equilibrium quantities rather than instantaneous values.

The uncertainty of the Hubble constant measurement using Eq. (\ref{eq:H0_direct}) is quantified as
\begin{equation}
\frac{\delta H_0}{H_0} \simeq 
\sqrt{\frac{1}{4}\left(\frac{\delta \langle F_{\rm obs} \rangle}{\langle F_{\rm obs} \rangle}\right)^2
+\frac{9}{4}\left(\frac{\delta \langle T_{\rm eff}^4 \rangle}{\langle T_{\rm eff}^4 \rangle}\right)^2
+4\left(\frac{\delta P_{\rm obs}}{P_{\rm obs}}\right)^2}.
\label{eq:dH/H}
\end{equation}
Since $\beta \ll 1$ is fixed by the hydrostatic-equilibrium structure of the envelope, its fractional uncertainty of $(1-\beta)$ is expected to be 
subdominant compared to observational errors and is neglected here. 
The $Q_0$ value may have uncertainty of a few percent, but its contribution to the $H_0$ uncertainty is currently sub-dominant as discussed latter.

An alternative $H_0$ formulation via surface gravity $g$ avoids assuming an Eddington ratio (Appendix \ref{appendix:logg}), but we prioritize the current formulation for our subsequent analysis due to lack of constraints on $\log g$.

\subsubsection{Reconstruction of Pulsation Periods}
\label{subsubsec:reconstruction_P}

Long-term time-domain monitoring in a decade is feasible for low-redshift LRDs.
For high-redshift sources, on the other hand, the observed period is further stretched by cosmic time dilation, $P_{\rm obs}=(1+z)P_{\rm src}$, 
making direct period measurements extremely expensive in terms of observational baseline.

An important exception arises when high-redshift LRDs are gravitationally lensed and produce multiple images. 
In such systems, the intrinsic pulsation signal is replicated in each image, but shifted in time due to gravitational lensing time delays and rescaled in flux by different lensing magnifications. 
By correcting both the time delays and the flux magnification, the intrinsic light curves can be reconstructed without requiring continuous monitoring over the full observed-frame period \citep[e.g., the \tgta\ and \tgtb\ in \rxc, where the observed-frame baseline reaches $\sim 130$ years;][]{Zhang_2025c}.

In the thin-lens approximation, the arrival time of a light ray observed at angular position {\mbox{\boldmath $\theta$}} is given by 
the lensing potential $\psi({\mbox{\boldmath $\theta$}})$ as
\begin{equation}
t({\mbox{\boldmath $\theta$}})=\frac{1+z_l}{c}\frac{D_{\rm ol} D_{\rm os}}{D_{\rm ls}}
\left[
\frac{1}{2}
|{\mbox{\boldmath $\theta$}}-{\mbox{\boldmath $\beta$}}_{\rm s}|^2
-\psi({\mbox{\boldmath $\theta$}})
\right],
\label{eq:fermat}
\end{equation}
where ${\mbox{\boldmath $\beta$}}_{\rm s}$ is the unlensed source position and $z_l$ is the redshift of the lens. $D_{\rm ol}$, $D_{\rm os}$, and $D_{\rm ls}$ are the angular diameter distances between the observer, lens, and source, respectively.
The expression in the bracket depends only on the lens mass model and is independent of cosmology.
Defining the time-delay distance as
\begin{equation}
D_{\Delta t}
\equiv
(1+z_l)\frac{D_{\rm ol} D_{\rm os}}{D_{\rm ls}}
\propto H_0^{-1},
\label{eq:deltatdist}
\end{equation}
the reconstructed pulsation period in the observer's frame scales as $P_{\rm obs}\propto D_{\Delta t}\propto H_0^{-1}$.
Thus, in lensed systems, the pulsation period itself becomes a cosmological observable through the time-delay distance.

Introducing the Fermat potential, $\tau \equiv |{\mbox{\boldmath $\theta$}}-{\mbox{\boldmath $\beta$}}_{\rm s}|^2/2-\psi({\mbox{\boldmath $\theta$}})$,
the fractional uncertainty in the Hubble constant can be expressed as
\begin{equation}
\frac{\delta H_0}{H_0} \simeq 
\sqrt{\frac{1}{4}\left(\frac{\delta \langle F_{\rm obs} \rangle}{\langle F_{\rm obs} \rangle}\right)^2
+\frac{9}{4}\left(\frac{\delta \langle T_{\rm eff}^4 \rangle}{\langle T_{\rm eff}^4 \rangle}\right)^2
+4\left(\frac{\delta \tau}{\tau}\right)^2}.
\label{eq:dH/H2}
\end{equation}
Note that $\delta \langle F_{\rm obs} \rangle / \langle F_{\rm obs} \rangle$ here, and in subsequent lensing cases, accounts for the magnification uncertainties derived from the lens mass model.
This expression clarifies that, for lensed cases, the dominant additional uncertainty arises from the reconstruction of the Fermat potential, i.e., the lens mass model, rather than from the intrinsic pulsation period itself.
Consequently, gravitational lensing provides a unique pathway to constrain the period even for sources at cosmological distances.

Multi-epoch monitoring of lensed images can extend the temporal baseline but introduces a mixture of cosmology-dependent time delays and cosmology-independent time intervals from the monitoring, therefore breaking the simple analytical relation. 
To resolve this complexity, we employ a maximum likelihood analysis to constrain $H_0$, with details provided in Appendix \ref{Appendix:likelihood}.
% Multiply imaged systems can also be monitored through multi-epoch observations to improve sampling and extend the temporal baseline.
% In practice, however, such monitoring introduces additional complexity. The time intervals between different lensed images depend inversely on $H_0$ through the time-delay distance, while the time intervals from the continuous monitoring of individual images solely track the rest-frame pulsation and do not depend on cosmology.
% With the mixture of these two types of time intervals, the simple analytical relation described above breaks down. 
% In this case, one can constrain $H_0$ using a maximum likelihood analysis, as described in Appendix \ref{Appendix:likelihood}.

\section{Observational Constrains and Future Prospects}
\label{sec:observation_constrain}

\subsection{Constraints from the Lensed LRD \tgta}
\label{subsec:first_constraint}

Strong gravitational lensing of LRDs by massive galaxy clusters provides a unique opportunity to study the variability of high-redshift LRDs over decades, leveraging time delays between multiple images. 
To date, four multiply imaged LRDs have been reported in the literature \citep[][]{Furtak_2023a,Baggen_2025,Golubchik_2025,Zhang_2025c}. 
Among these, \tgta, recently identified by \citet{Zhang_2025c}, is uniquely the only known LRD exhibiting clear pulsation signatures.
% With its predicted time delay of more than one hundred years, \tgta\ provides an unparalleled opportunity to sample the \hbox{full-cycle} light curve and constrain the pulsation period. We use this object to demonstrate the cosmological inference framework proposed by this work.
With a predicted time delay exceeding a century, \tgta\ allows for full-cycle light curve sampling to constrain the pulsation period. We use this object to demonstrate our cosmological inference framework.

% To propagate uncertainties from the observable parameters into $H_0$, we employ a Monte Carlo approach. 
We adopt a Monte Carlo (MC) approach to propagate the uncertainties from the observable parameters, $P_{\rm src}$, $T_{\rm eff}$, and $F_{\rm obs}$, into $H_0$. 
The source pulsation period is determined as $P_{\rm src} = 40.8^{+6.7}_{-8.9}$ yr through multi-band sinusoidal light-curve modeling following \citet{Zhang_2025c}. The effective temperature and bolometric flux are derived by fitting the continuum SED with a blackbody plus power-law model, yielding $\langle T_{\rm eff} \rangle = 3834^{+77}_{-80}~\rm K$ and $\langle F_{\rm obs}\rangle = 8.87^{+0.25}_{-0.22}\times10^{-16}~\rm erg~s^{-1}~cm^2$. Currently, the $\sim 25\%$ uncertainty in $P_{\rm src}$ dominates the total error budget, primarily due to the sparse sampling of the light curves. Detailed descriptions of the Bayesian inference and prior constraints are provided in Appendix \ref{appendix:lc_sed}.

To isolate the $H_0$ dependence in $P_{\rm obs}$, we define $P_{\rm obs} = \mathcal{P}_z H_0^{-1}$, where $\mathcal{P}_z$ a quantity primarily determined by the lens model. Then we obtain
\begin{equation}
\begin{split}
H_0 = \frac{\kappa_{\rm es}}{c^2(1-\beta)(1+z)^3 }\frac{ \langle \sigma_{\rm SB} T_{\rm eff}^4 \rangle^{3/2}}{\langle F_{\rm obs} \rangle^{1/2}} \\
      \times \left(\frac{\mathcal{P}_z }{2\pi Q_0}\right)^{2} \left(\int_{0}^{z}\frac{dz'}{E(z')}\right)^{-1}.
\end{split}
\label{eq:H0_lensing_case}
\end{equation}
While $\mathcal{P}_z$ retains a minor redshift dependence through the ratio $D_{\rm os}/D_{\rm ls}$, this effect is negligible when the source redshift significantly exceeds that of the lens. We derive the posterior distribution of $\mathcal{P}_z$ by multiplying $P_{\rm obs}$ by the fiducial $H_0$ used in the lens model.

Finally, we obtain an estimation of $H_0 = 120.7^{+47.0}_{-46.5}\ \mathrm{km\,s^{-1}\,Mpc^{-1}}$. Although the central value is high, it remains consistent with local measurements within the $1\sigma$ uncertainty ($\sim 40\%$), which is primarily driven by the rest-frame period constraint.
Despite the current limited precision, this result serves as a critical proof of concept for using LRD pulsations as an independent cosmological probe.

\subsection{Physical Validation and Systematic Uncertainties}
\label{subsec:physical_validation}

Future JWST multi-cycle spectrophotometric monitoring will be essential to validate the pulsational nature of LRDs and the $P$-$L$-$T_{\rm eff}$ relation, establishing LRD pulsations as a robust cosmological probe. 
% Beyond merely reducing statistical uncertainties on $P_{\rm src}$, these observations will verify the underlying pulsating nature of LRDs and validate the predicted $P$-$L$-$T_{\rm eff}$ relation.  
% Specifically, the pulsational nature predicts a periodic expansion and contraction of the envelope, which should manifest as detectable velocity shifts ($\sim 100\text{ km s}^{-1}$) in absorption lines (e.g., H$\alpha$, H$\beta$, He I$\lambda$10830). Such spectroscopic signatures, analogous to the atmospheric dynamics of pulsating stars, would provide independent kinematic evidence to complement the periodic photometric light curves \citep[][]{Zhang_2025c}. 
Beyond periodic light curves, the pulsation predicts periodic envelope expansion and contraction, which should manifesting as $\sim 100\text{ km s}^{-1}$ velocity shifts in absorption lines (e.g., H$\alpha$, He I$\lambda$10830). Such spectroscopic signatures, analogous to the atmospheric dynamics of pulsating stars, would provide kinematic evidence independent of photometry \citep{Zhang_2025c}.
% The discovery and monitoring of additional gravitationally lensed LRDs low-redshift LRDs will be essential to test the theoretical $P$-$L$-$T_{\rm eff}$ relation. 
% If the predicted physical scaling is observationally confirmed across a statistically significant sample, it will firmly anchor the $P$-$L$-$T_{\rm eff}$ relation, positioning LRDs as a reliable and independent probe for $H_0$.
Furthermore, expanding the sample with additional lensed or low-redshift LRDs will allow empirical calibration of the relation, anchoring LRDs as a reliable $H_0$ probe.

\begin{figure*}
\centering
\includegraphics[width=0.95\textwidth]{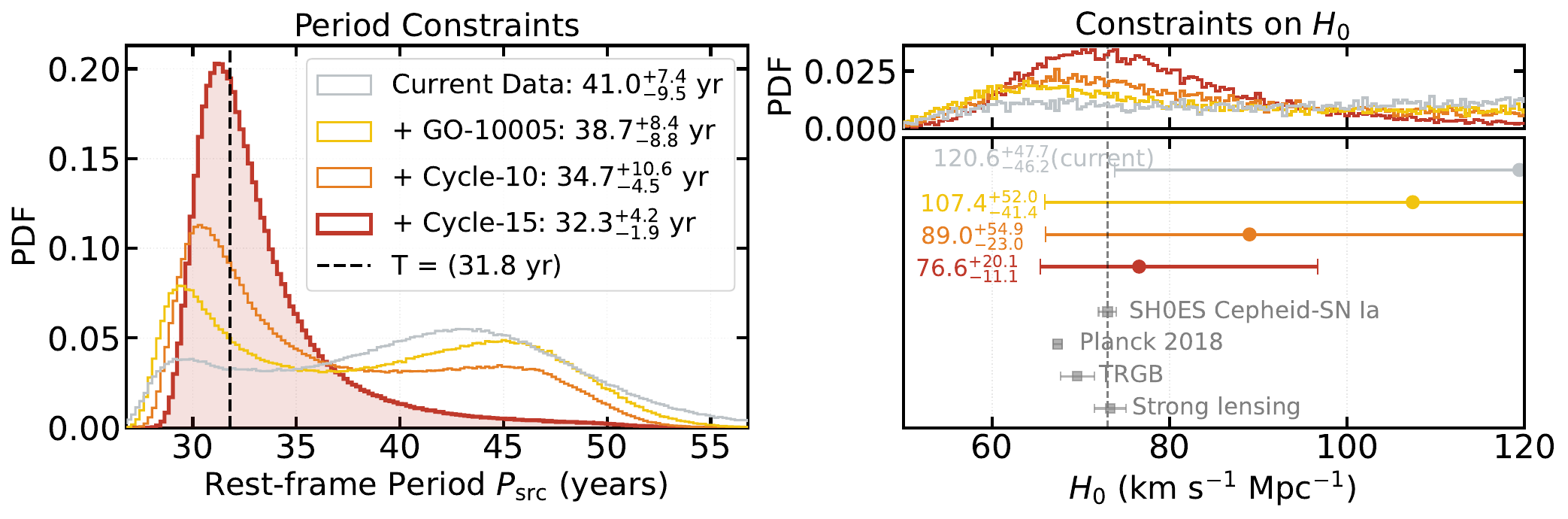}
\caption{Left: Posterior probability density functions (PDF) for the rest-frame period $P_{\rm src}$. 
% These constraints include predictions for the approved JWST program GO-10005 upon completion, as well as extended monitoring scenarios: (i) program 10005 combined with an additional program conducted 5 years later (JWST Cycle 10), and (ii) program 10005 combined with two additional programs conducted 5 and 10 years later (JWST Cycles 10 and 15).
We consider the approved program GO-10005 and two extended scenarios incorporating additional monitoring over 5-year and 10-year baselines.
The uncertainty $\sigma_{P}$ diminishes with extended monitoring, converging toward the assumed value of 31.8 years.
Right: The corresponding constraints on the Hubble constant $H_0$. Top panel shows the aggregated PDF of $H_0$ derived from different period measurements (correspond to the color in the left panel), showing the significant decrease in uncertainties as monitoring accumulated.
Bottom panel shows comparison of $H_0$ measurement precision across different scenarios. The colored points represent our simulated results with $1\sigma$ uncertainties. Existing constraints (gray squares) including SH0ES \citep[][]{Riess_2022_SHOES}, Planck \citep[][]{Planck_2020}, TRGB \citep[][]{Freedman_2020_TRGB}, and Strong Lensing \citep[][]{Wong_2020_lensingH0} are shown for comparison. The vertical dashed line $H_0 = 73\text{ km s}^{-1}\text{Mpc}^{-1}$.
}
\label{fig:H0_constrain}
% \vspace{2mm}
\end{figure*}

Assuming the physical scaling holds, the precision of this cosmological framework is governed by the pulsation period $P_{\rm src}$ determination and the potential systematic departures from the idealized $P$-$L$-$T_{\rm eff}$ relation.  
The current theoretical framework assumes specific values for the $Q_{0}$ and $\beta$. 
% Their systematic error budget would be:
% \begin{equation}
% \frac{\delta H_0}{H_0} \simeq 
% \sqrt{4\left(\frac{\delta Q_0}{Q_0} \right)^2
% +\left(\frac{\delta \beta}{1-\beta}\right)^2}.
% \label{eq:dH/H}
% \end{equation}
{Under the self-consistent assumption that the gas envelope is supported by radiation pressure, $\beta \ll 1$ is naturally constrained, minimizing its error contribution. 
% If we assign a conservative theoretical uncertainty of $\sim 5\%$ to the pulsation constant $Q_0$ (reflecting variations in envelope geometry or polytropic index), the resulting systematic floor for $H_0$ would be approximately $10\%$.
Assigning a conservative $5\%$ theoretical uncertainty (reflecting variations in envelope geometry or polytropic index) to the pulsation constant $Q_0$ yields an $H_0$ systematic floor of $\sim 10\%$.}
Currently, the $\sim 25\%$ uncertainty in $P_{\rm src}$ dominates the error budget.
% It is only after 5–10 years of high-cadence and long-term monitoring—when the period is constrained to high precision—that the uncertainty floor will become dominated by the aforementioned systematic effects. 
The systematic floor will only become limiting after $\gtrsim 5$ years of monitoring constrains the period to high precision (Fig.~\ref{fig:H0_constrain}).
Future detailed modeling and empirical calibration using a larger sample of lensed and/or low-redshift LRDs will be essential to precisely characterize these systematic terms. 
% In particular, low-redshift LRDs offers a unique laboratory.
% These sources can provide high signal-to-noise data and the observational feasibility for the high-cadence, long-term monitoring required to empirically calibrate parameters such as the pulsation constant $Q_0$ and the Eddington ratio dependence. 
In particular, low-redshift LRDs offer a unique laboratory, providing both high signal-to-noise data and the observational feasibility for the high-cadence, long-term monitoring required to empirically calibrate parameters such as $Q_0$.
A well-calibrated local framework will significantly reduce the systematic floor when applying this method to the high-redshift population discovered by JWST.

Additional uncertainties stem from $T_{\rm eff}$ and $F_{\rm obs}$, which will only become a limitation once $P_{\rm src}$ is precisely constrained and the systematic uncertainties of the $P$-$L$-$T_{\rm eff}$ are well understood. 
Current analysis for \tgta\ relies only on JWST/NIRCam photometry, where the optical SED can suffers from a degeneracy between $T_{\rm eff}$, dust extinction, and the Balmer break strength, potentially introducing systematic biases \citep[][]{Liu2026}.
% Future observations in more wavelengths will be important in breaking these degeneracies. 
% Specifically, NIRSpec/prism spectroscopy will better determine the continuum shape and allow more reliable constrain on the Balmer break strength. Mid-infrared photometry from MIRI will probe the \hbox{long-wavelength} slope of the SED, effectively isolating the effect of dust extinction.
Future multiwavelength observations will help break these degeneracies: JWST/NIRSpec prism spectroscopy can better constrain the continuum and Balmer break strength, while JWST/MIRI photometry can isolate dust extinction through the long-wavelength SED slope.

Finally, for lensed sources, the inferred $H_0$ is subject to the mass-sheet degeneracy \citep{Falco_1985}. An external convergence $\kappa_{\rm ext}$ scales the observed flux as $F_{\rm obs} \propto (1-\kappa_{\rm ext})^2$ and the reconstructed pulsation period as $P_{\rm obs} \propto (1-\kappa_{\rm ext})^{-1}$ \citep[][]{Suyu_2013}. Consequently, the $H_0$ derived from Eq. (\ref{eq:H0_lensing_case}) would be affected as $H_0 \propto (1-\kappa_{\rm ext})$. This systematic effect can be effectively quantified and mitigated by characterizing the lens environment and line-of-sight mass distribution using high-resolution imaging \citep[][]{Greene_2013}.

\subsection{Forecast for $H_0$ Constraints}
\label{subsec:forecast}
We now assess the achievable constraints on $H_0$ from this framework once it is validated, assuming continuous monitoring of \tgta\ over the next 10 years and a true $P_{\rm src} = 31.8$ years. 
We ignore the complexity due to the mixture of lensing time delays and direct monitoring here.
We consider three epochs: the completion of Cycles 5/6 JWST program GO-10005 (+1, +1.5, +2, and +2.5 yr baseline), and extended monitoring at 5 and 10 years (e.g., Cycles 10 and 15).
For each configuration, we perform 50 independent realizations, incorporating random Gaussian noise into the mock datasets to match current measurement uncertainties.

As shown in Fig. \ref{fig:H0_constrain} (left), the posterior distribution for the rest-frame period $P_{\rm src}$ sharpens significantly as the monitoring baseline extends. 
% While the initial measurement (gray) exhibits large uncertainty and a significant deviation from the true value, five years of monitoring (orange) reduce the uncertainty to below five years. With a 10-year monitoring baseline (purple), we can achieve precision of a few years in $P_{\rm src}$.
The uncertainty reduces to $<5$ yr after 5 years of monitoring, even reaching sub-year precision for shorter intrinsic periods (see Appendix~\ref{Appendix:period}).
% We assume that the uncertainty in $Q_0$ can be constrained to $\sim3\%$ and the systematic uncertainties in $F_{\rm obs}$ and $T_{\rm eff}$ discussed in Section \ref{subsec:physical_validation} can be effectively mitigated through extensive observations, such that only statistical uncertainties of $\sim1\%$ remain.
% The impact on cosmological constraints is shown in Fig. \ref{fig:H0_constrain} (right). 
Assuming a $3\%$ uncertainty in the structure constant $Q_0$ and mitigated systematics ($\sim 1\%$ statistical noise) in $F_{\rm obs}$ and $T_{\rm eff}$ through extensive observations, we project the cosmological impact in Fig.~\ref{fig:H0_constrain} (right).
The upper panel illustrates the expected convergence of the $H_0$ PDF toward the fiducial value. In the lower panel, we compare the precision against standard benchmarks from SH0ES \citep[][]{Riess_2022_SHOES}, Planck \citep[][]{Planck_2020}, tip of the red giant branch \citep[TRGB method;][]{Freedman_2020_TRGB}, and strong gravitational lensing \citep[][]{Wong_2020_lensingH0}.

Our forecast indicates that with an extended monitoring campaign, a single multiply imaged LRD can achieve a competitive measurement of $H_0$ with an uncertainty of $\sim 15\%$, potentially reaching a uncertainty level of $\lesssim 5\%$ in the most optimistic case (intrinsically short period). This result indicates that long-term time-domain monitoring is the key observational requirement for this method.

Additionally, in the future, if a sample of $N_{\rm LRD}^{\rm var}$ LRDs exhibiting measurable pulsations becomes available, the uncertainty is expected to decrease as
\begin{align}
\frac{\delta H_0}{H_0} &\simeq 
\frac{2}{\sqrt{N_{\rm LRD}^{\rm var}}} 
\frac{\delta P_{\rm obs}}{P_{\rm obs}}\\ \nonumber
&\sim 
6.3\% 
\left(\frac{N_{\rm LRD}^{\rm var}}{10}\right)^{-1/2}\left(\frac{\delta P_{\rm obs}/P_{\rm obs}}{0.1}\right).
\label{eq:multi_lrd}
\end{align}
Therefore, even a modest sample of $N_{\rm LRD}^{\rm var}\gtrsim4$ well-characterized pulsating LRDs would allow constraints at the $\lesssim10\%$ level.

\section{Summary}
\label{sec:summary_discussion}

% In this work, we propose a novel cosmological framework that connects the quasi-periodic variability of LRDs to cosmological distance measurements. 
% By linking the pulsation period to the global properties of radiation-pressure-supported envelopes, we derive an idealized $P$--$L$--$T_{\rm eff}$ relation that enables a direct inference of the Hubble constant.
In this work, we propose a novel cosmological framework linking the quasi-periodic variability of LRDs to distance measurements via a physically motivated $P$--$L$--$T_{\rm eff}$ relation for idealized radiation-pressure-supported envelopes.
Applying this framework to the lensed LRD R2211-RX1, we obtain a preliminary constraint of $H_0 = 120.7^{+47.0}_{-46.5}\ \mathrm{km\ s^{-1}\ Mpc^{-1}}$. Although the uncertainty remains large, it is dominated by the limited constraint on the pulsation period and serves as a critical proof-of-concept. Our forecast indicates that 10 years of monitoring can reduce $H_0$ statistical uncertainty to 3--20\%, depending on the intrinsic period.

The ultimate accuracy of this framework depends on quantifying systematic uncertainties in the $P$--$L$--$T_{\rm eff}$ relation, which remain to be fully characterized. 
Empirical calibration via low-redshift LRDs and improved modeling are essential to mitigate these effects and validate the theoretical baseline. 
Once calibrated, this method will provide a robust, independent $H_0$ probe.
Though nascent, this approach offers a complementary test for cosmological models, particularly at high redshifts where standard candles are scarce.
% Although in its nascent stages, such an approach provides a complementary avenue to test cosmological models at an epoch that remains largely challenging for standard distance indicators.
% Future observations for a modest sample of pulsating LRDs are essential to empirically validate and calibrate this physically motivated theoretical baseline, decreasing the systematic uncertainties
% The ultimate accuracy of this framework will likely be limited by systematic uncertainties associated with the $P$--$L$--$T_{\rm eff}$ relation. 
% These systematics remain to be fully quantified.
% Empirical calibration using low-redshift LRDs and improved theoretical modeling will be essential to control these effects.
% it will provide a robust, independent probe to constrain the Hubble constant, particularly at high redshifts where standard candles like Type Ia supernovae are scarce. 

% In summary, LRD pulsations offer a promising and conceptually distinct route to probe cosmology at high redshift. 

\begin{acknowledgments}
We acknowledge support from the National Natural Science Foundation of China (12225301, 12573015, W2532003), the Beijing Natural Science Foundation (IS25003), and the China Manned Space Program (CMS-CSST-2025-A09).
M.O. acknowledges support from JSPS KAKENHI Grant Numbers JP25H00662 and JP22K21349.
We thank the JWST VENUS team, led by Seiji Fujimoto and Dan Coe, for the execution of the program (GO-6882) and for their commitment to making the data publicly available.
We thank Subo Dong and Cheng Zhao for helpful discussions.
\end{acknowledgments}

\appendix
% \restartappendixnumbering

\section{Likelihood Analysis Framework for Multi-epoch Observations of Lensed Systems}
\label{Appendix:likelihood}

To self-consistently constrain $H_0$ for cases with a mixture of $H_0$-dependent (lensing-derived) and $H_0$-independent (monitoring-derived) timescales, we can employ a maximum likelihood analysis. 
This approach treats the observed data from all multiple images as a single realization of the source's intrinsic light curve, transformed by the lens. 
We define a likelihood function $\mathcal{L}(H_0, \vec{\theta})$ that incorporates both the lens model parameters and the pulsation model:

% \begin{equation}
%     \ln \mathcal{L}(H_0, \vec{\theta}) = -\frac{1}{2} \sum_{j=1}^{N_{\rm img}} \sum_{i=1}^{N_{\rm obs, j}} \left[ \frac{\left( F_{i,j}^{\rm obs} - \mu_j F_{\rm model}(t_{i} - \Delta t_j(H_0); \vec{\theta}_{\rm var}) \right)^2}{\sigma_{i,j}^2} \right] + \ln \mathcal{P}(\vec{\theta})
% \end{equation}
\begin{widetext}
\begin{equation}
    \ln \mathcal{L}(H_0, \vec{\theta}) = -\frac{1}{2} \sum_{j=1}^{N_{\rm img}} \sum_{i=1}^{N_{\rm obs, j}} \left[ \frac{\left( F_{i,j}^{\rm obs} - \mu_j F_{\rm model}(t_{i} - \Delta t_j(H_0); \vec{\theta}_{\rm var}) \right)^2}{\sigma_{i,j}^2} \right] + \ln \mathcal{P}(\vec{\theta})
\end{equation}
\end{widetext}
Each term in this expression is defined as follows: 
\begin{itemize}
    \item $F_{i,j}^{\rm obs}$ and $\sigma_{i,j}$: The observed flux and its associated uncertainty for the $i$-th observation of the $j$-th lensed image at observed time $t_{i}$.
    \item $\mu_j$: The magnification factor for image $j$. These are treated as nuisance parameters constrained by the lens mass model.
    \item $\Delta t_j(H_0)$: The relative time delay for image $j$ with respect to the reference image. Crucially, the cosmology dependence can be explicitly isolated in the time delay $\Delta t_j(H_0) = \frac{D_{\Delta t}(H_0)}{c} \Delta \tau_j$, where $\Delta \tau_j$ is the Fermat potential difference relative to the leading image. $\Delta \tau_j$ are treated as nuisance parameters
    \item $F_{\rm model}$: The template for the intrinsic variability, such as the simple pulsation model $F_{\rm model}(t) = f_0 [1 + a \sin(2\pi t / P_{\rm src} + \phi)]$. 
    \item $\vec{\theta}$: The set of all parameters (except $H_0$) to be explored, which can be subdivided into $\vec{\theta} = \{ \vec{\theta}_{\rm var}, \vec{\theta}_{\rm lens}\}$:  
    \begin{itemize}
        \item $\vec{\theta}_{\rm var} = \{P_{\rm src}, f_0, a, \phi, \rm etc\}$: Parameters describing the intrinsic pulsation, where $P_{\rm src}$ is the rest-frame period.
        \item $\vec{\theta}_{\rm lens}= \{\Delta \tau_j, \mu_j\}$: Parameters from the lens mass model, which serve as nuisance parameters and carry their own systematic uncertainties. 
    \end{itemize}
    \item  $\ln \mathcal{P}(\vec{\theta})$: The prior distribution, which incorporates independent constraints from the lens model  and theoretical limits on LRD pulsation parameters (e.g., $Q_0$). 
\end{itemize}

The full parameter vector $\vec{\theta}$ encompasses not only the intrinsic LRD variability properties ($\vec{\theta}_{\rm var}$) but also the lens model parameters ($\mu_j$ and $\Delta \tau_j$). The second term, $\ln \mathcal{P}(\vec{\theta})$, serves as an informative prior that links the light curve modeling to the lens model. Specifically, the joint posterior distributions of the magnifications and Fermat potentials derived independently from the lens mass model are adopted as the priors $\mathcal{P}(\mu_j, \Delta \tau_j)$. By exploring this full high-dimensional parameter space using MCMC methods, we effectively marginalize over the nuisance parameters inherited from the lens mass model. This approach ensures that the systemic uncertainties of both the magnification and the time delays are robustly propagated into the final posterior constraint on $H_0$.

\section{Alternative $H_0$ Formulation via Surface Gravity}
\label{appendix:logg}
From Eqs. (\ref{eq:pulsation})--(\ref{eq:sb}), one can also obtain:
\begin{equation}
    (GM)^{1/3} = \frac{\kappa_{\rm es}\sigma_{\rm SB} T_{\rm eff}^4}{c(1-\beta)}\left(\frac{P_{\rm src}}{Q}\right)^{4/3} = g\left(\frac{P_{\rm src}}{Q}\right)^{4/3},
\end{equation}
where $g(=GM/R^2)$ is the surface gravity of the photosphere.
This gives the second formulation for the Hubble constant:
\begin{equation}
H_0 = \frac{cQ^2(1+z)^3}{gP_{\rm obs}^2} \cdot \frac{\langle F_{\rm obs} \rangle^{1/2}}{\langle \sigma_{\rm SB} T_{\rm eff}^4 \rangle^{1/2}} \int_0^z \frac{dz'}{E(z')}.
\label{eq:H0_gravity}
\end{equation}
This expression avoids assuming an Eddington ratio or direct electron scattering opacity by instead linking the physical scale of the photosphere to surface gravity $g$. 
While empirical calibrations using features like the CaT equivalent width \citep[EW;][]{Cenarro2002} can estimate $\log g$, significant uncertainties remain regarding metallicity degeneracy and the applicability of stellar models to LRDs. This limitation may be mitigated with improved LRD atmosphere modeling and more robust spectroscopic diagnostics \citep[][]{Liu2026}.

\begin{figure*}
\centering
\includegraphics[width=0.9\textwidth]{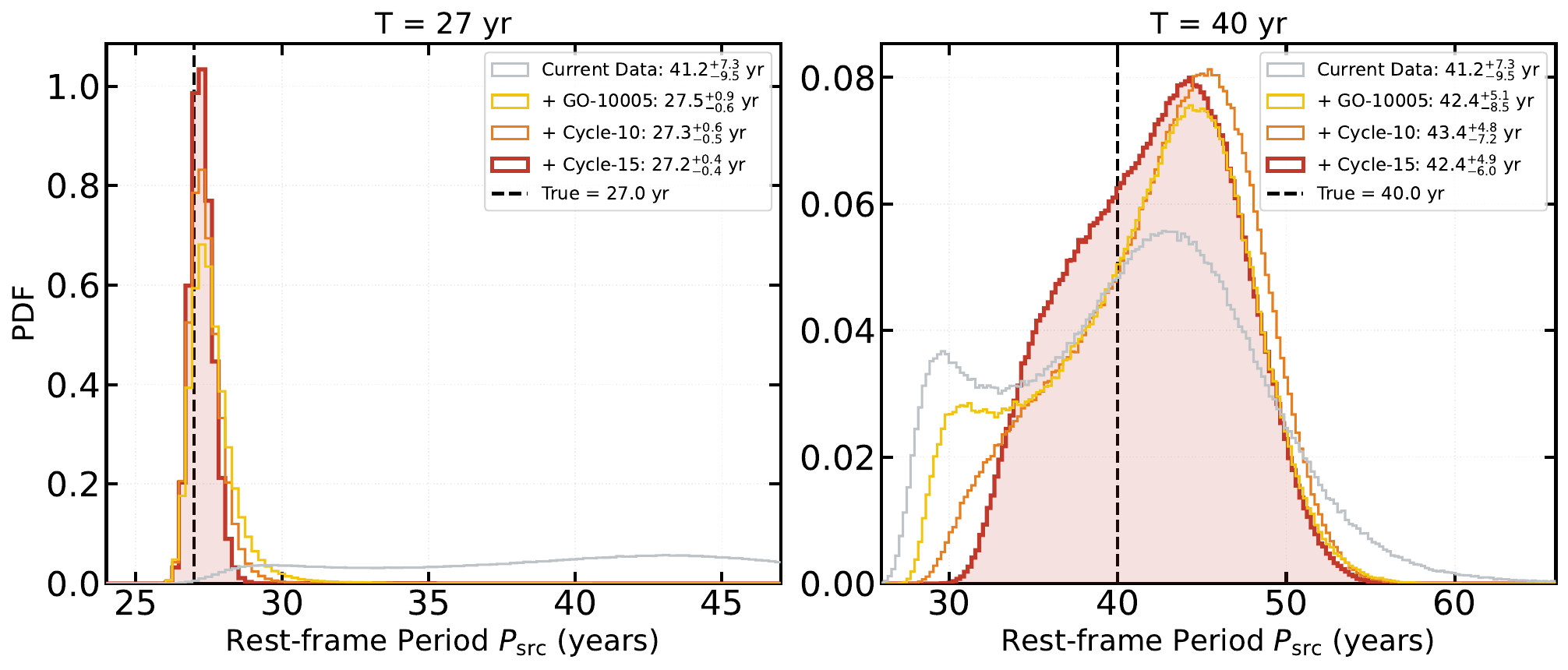}
\caption{Posterior probability density functions for the rest-frame period $P_{\rm src}$ with intrinsic $P_{\rm src}$ of 27 years (left) and 40 years (right). The additional monitoring configuration is the same as that in Fig. \ref{fig:H0_constrain}.}
\label{fig:period_constrain}
\vspace{2mm}
\end{figure*}

\section{Multi-band Light-curve Modeling and SED Fitting}
\label{appendix:lc_sed}
% To reconstruct the observed pulsation period $P_{\rm src}$, we follow \citet{Zhang_2025c} to perform a joint multi-band modeling of the rest-frame light curves of \tgta with a sinusoidal pulsation: 
Following \citet{Zhang_2025c}, we model the multi-band rest-frame light curves with a sinusoidal pulsation: $F_\lambda(t) = f_{0,\lambda} [1 + a_\lambda \sin(2\pi t / P_{\rm src} + \phi)]$, assuming $z=4.3$ and the fiducial lens model \citep{Zhang_2025c}.
The redshift will be precisely determined by future spectroscopic observations.
% The pulsation period $P_{\rm src}$ and the phase $\phi$ are global parameters shared across all bands, while the baseline flux $f_{0,\lambda}$ and the relative amplitude $a_\lambda$ are allowed to vary for each filter to account for the SED and wavelength-dependent variability amplitudes.
The period $P_{\rm src}$ and phase $\phi$ are shared across all bands, while $f_{0,\lambda}$ and relative amplitudes $a_\lambda$ are allowed to vary by filter.
% We employed a Bayesian framework to estimate the posterior distributions of these parameters using emcee \citep{Foreman-Mackey_2013}. 
% To ensure physical plausibility, we imposed uniform priors on the baseline fluxes ($f_{0,\lambda} > 0$) and constrained the fractional semi-amplitude $a_\lambda$ to be less than 50\%. 
We estimate posteriors using \texttt{emcee} \citep{Foreman-Mackey_2013}, with uniform priors $f_{0,\lambda} > 0$ and $a_\lambda < 50\%$. 
% This constraint is consistent with the pulsational model of \citet{Cantiello_2025}, where non-linear oscillations driven by the $\kappa$-mechanism in quasi-star envelopes are expected to saturate or trigger enhanced mass-loss (super-winds) before reaching extreme amplitudes \citep[see also][]{Appenzeller_1970,Papaloizou_1973}.
% Practically, this limit prevents unphysical, degenerate solutions where a sparsely sampled light curve is fitted as a small segment of a massive-amplitude oscillation with a multi-century period, thereby focusing the analysis on the decadal modes theoretically predicted for first overtone population.
This amplitude limit aligns with pulsational models where non-linear oscillations saturate before reaching extreme amplitudes \citep{Cantiello_2025, Papaloizou_1973}. Practically, this constraint prevents unphysical, long-period degeneracies arising from sparse sampling, focusing the fit on theoretically predicted decadal modes.
% The walkers for the period parameter were uniformly distributed across a broad prior range of $[2, 200]$ years.
% We finally obtain $P_{\rm src} = 40.8^{+6.7}_{-8.9}$ yr, the current uncertainty is therefore $\sim 25\%$. 
With a broad prior on $P_{\rm src} \in [2, 200]$ yr, we obtain $P_{\rm src} = 40.8^{+6.7}_{-8.9}$ yr.
The uncertainty in $P_{\rm src}$ accounts for both the reconstruction of the Fermat potential (i.e., the uncertainties of the time delay and magnification from the lens mass model) and the fitting error arising from insufficient sampling. 
At present, the latter dominates the total uncertainty budget.

For $T_{\rm eff}$ and $F_{\rm obs}$, we model the continuum emission of \tgta\ at each epoch as a superposition of a blackbody component and a fixed underlying power-law component, as in \cite{Zhang_2025c}.
We obtain the covariance between $T_{\rm eff}$ and $F_{\rm obs}$ at each epoch using emcee.
Averaging posterior samples across four epochs yields $\langle T_{\rm eff} \rangle = 3834^{+77}_{-80}~\rm K$ and $\langle F_{\rm obs}\rangle = 8.87^{+0.25}_{-0.22}\times10^{-16}~\rm erg~s^{-1}~cm^2$.
% The small uncertainties arise because the averaged quantities are jointly constrained by posterior samples from multiple epochs rather than a single-epoch measurement, although the result is partially model-dependent. Nevertheless, the uncertainties in $T_{\rm eff}$ and $F_{\rm obs}$ remain much smaller than the current uncertainty in $P_{\rm src}$.
These $\sim 2\%$ uncertainties are significantly smaller than that of $P_{\rm src}$, though they remain partially model-dependent.

% Since $P_{\rm obs}$ (and consequently $P_{\rm src}$) depends on the time-delay reconstruction (Eq. \ref{eq:fermat}), it introduces a dependence on $H_0$. 
% To isolate this, we define $P_{\rm obs} = \mathcal{P}_z H_0^{-1}$, where $\mathcal{P}_z$ is a quantity primarily determined by the lens model. Then we obtain

\section{Period constrains for different intrinsic periods}
\label{Appendix:period}
We also perform the forecast simulation assuming true periods of 27 and 40 years. Fig. \ref{fig:period_constrain} shows the forecast PDFs. The constraint on the period is strongly dependent on the intrinsic timescale, with shorter periods yielding tighter constraints and, in some cases, reaching sub-year precision.

\bibliography{ref}% Produces the bibliography via BibTeX.

@ARTICLE{Liu2026,
       author = {{Liu}, Hanpu and {Jiang}, Yan-Fei and {Quataert}, Eliot and {Greene}, Jenny E. and {Ma}, Yilun and {Lin}, Xiaojing},
        title = "{Synthetic Spectral Library of Optically Thick Atmospheres for Little Red Dots}",
      journal = {arXiv e-prints},
         year = 2026,
        month = mar,
          eid = {arXiv:2603.02317},
        pages = {arXiv:2603.02317},
          doi = {10.48550/arXiv.2603.02317},
archivePrefix = {arXiv},
       eprint = {2603.02317},
 primaryClass = {astro-ph.GA},
       adsurl = {https://ui.adsabs.harvard.edu/abs/2026arXiv260302317L}
}

@ARTICLE{Papaloizou_1973,
       author = {{Papaloizou}, J.~C.~B.},
        title = "{Non-linear pulsations of upper main sequence stars-I.A perturbation approach}",
      journal = {Monthly Notices of the Royal Astronomical Society},
         year = 1973,
        month = jan,
       volume = {162},
        pages = {143},
          doi = {10.1093/mnras/162.2.143},
       adsurl = {https://ui.adsabs.harvard.edu/abs/1973MNRAS.162..143P}
}

@ARTICLE{Lin_2026,
       author = {{Lin}, Xiaojing and {Fan}, Xiaohui and {Cai}, Zheng and {Bian}, Fuyan and {Liu}, Hanpu and {Sun}, Fengwu and {Ma}, Yilun and {Greene}, Jenny E. and {Strauss}, Michael A. and {Green}, Richard and {Lyu}, Jianwei and {Champagne}, Jaclyn B. and {Goulding}, Andy D. and {Inayoshi}, Kohei and {Jin}, Xiangyu and {Leung}, Gene C.~K. and {Li}, Mingyu and {Liu}, Weizhe and {Liu}, Yichen and {Mao}, Junjie and {Pudoka}, Maria Anne and {Tee}, Wei Leong and {Wang}, Ben and {Wang}, Feige and {Wu}, Yunjing and {Yang}, Jinyi and {Zhang}, Haowen and {Zhu}, Yongda},
        title = "{The Discovery of Little Red Dots in the Local Universe: Signatures of Cool Gas Envelopes}",
      journal = {The Astrophysical Journal},
         year = 2026,
        month = feb,
       volume = {997},
       number = {2},
          eid = {364},
        pages = {364},
          doi = {10.3847/1538-4357/ae2bdf},
archivePrefix = {arXiv},
       eprint = {2507.10659},
 primaryClass = {astro-ph.GA},
       adsurl = {https://ui.adsabs.harvard.edu/abs/2026ApJ...997..364L}
}

@ARTICLE{Golubchik_2025,
       author = {{Golubchik}, Miriam and {Furtak}, Lukas J. and {Allingham}, Joseph F.~V. and {Zitrin}, Adi and {Akins}, Hollis B. and {Kokorev}, Vasily and {Fujimoto}, Seiji and {Abdurro'uf} and {Amor{\'\i}n}, Ricardo O. and {Bauer}, Franz E. and {Bezanson}, Rachel and {Brada{\v{c}}}, Marusa and {Bradley}, Larry D. and {Brammer}, Gabriel B. and {Chisholm}, John and {Coe}, Dan and {Conselice}, Christopher J. and {Dayal}, Pratika and {Dessauges-Zavadsky}, Miroslava and {Diego}, Jose M. and {Faisst}, Andreas L. and {Fei}, Qinyue and {Ferguson}, Henry C. and {Finkelstein}, Steven L. and {Frye}, Brenda L. and {Gonz{\'a}lez-Otero}, Mauro and {Greene}, Jenny E. and {Harikane}, Yuichi and {Hsiao}, Tiger Yu-Yang and {Inayoshi}, Kohei and {Jim{\'e}nez-Teja}, Yolanda and {Knudsen}, Kirsten and {Koekemoer}, Anton M. and {Labb{\'e}}, Ivo and {Lucas}, Ray A. and {Magdis}, Georgios E. and {Matthee}, Jorryt and {Messa}, Matteo and {Naidu}, Rohan P. and {Nakane}, Minami and {Noirot}, Ga{\"e}l and {Pan}, Richard and {Papovich}, Casey and {Richard}, Johan and {Ricotti}, Massimo and {Robbins}, Luke and {Stark}, Daniel P. and {Sun}, Fengwu and {Treu}, Tommaso and {Tripodi}, Roberta and {Vanzella}, Eros and {Willott}, Chris and {Windhorst}, Rogier A.},
        title = "{VENUS: When Red meets Blue -- A multiply imaged Little Red Dot with an apparent blue companion behind the galaxy cluster Abell 383}",
      journal = {arXiv e-prints},
         year = 2025,
        month = dec,
          eid = {arXiv:2512.02117},
        pages = {arXiv:2512.02117},
archivePrefix = {arXiv},
       eprint = {2512.02117},
 primaryClass = {astro-ph.GA},
       adsurl = {https://ui.adsabs.harvard.edu/abs/2025arXiv251202117G}
}

@ARTICLE{Inayoshi_2026b,
       author = {{Inayoshi}, Kohei and {Murase}, Kohta and {Kashiyama}, Kazumi},
        title = "{Spectral Uniformity of Little Red Dots: A Natural Outcome of Coevolving Seed Black Holes and Nascent Starbursts}",
      journal = {The Astrophysical Journal},
         year = 2026,
        month = mar,
       volume = {1000},
       number = {1},
          eid = {90},
        pages = {90},
          doi = {10.3847/1538-4357/ae42ce},
archivePrefix = {arXiv},
       eprint = {2509.19422},
 primaryClass = {astro-ph.GA},
       adsurl = {https://ui.adsabs.harvard.edu/abs/2026ApJ..1000...90I}
}

@ARTICLE{Inayoshi_2025a,
       author = {{Inayoshi}, Kohei},
        title = "{Little Red Dots as the Very First Activity of Black Hole Growth}",
      journal = {The Astrophysical Journal Letter},
         year = 2025,
        month = jul,
       volume = {988},
       number = {1},
          eid = {L22},
        pages = {L22},
          doi = {10.3847/2041-8213/adea66},
archivePrefix = {arXiv},
       eprint = {2503.05537},
 primaryClass = {astro-ph.GA},
       adsurl = {https://ui.adsabs.harvard.edu/abs/2025ApJ...988L..22I}
}

@ARTICLE{Zhang_2025,
       author = {{Zhang}, Zijian and {Jiang}, Linhua and {Liu}, Weiyang and {Ho}, Luis C.},
        title = "{Analysis of Multi-epoch JWST Images of {\ensuremath{\sim}}300 Little Red Dots: Tentative Detection of Variability in a Minority of Sources}",
      journal = {The Astrophysical Journal},
         year = 2025,
        month = may,
       volume = {985},
       number = {1},
          eid = {119},
        pages = {119},
          doi = {10.3847/1538-4357/adcb3e},
archivePrefix = {arXiv},
       eprint = {2411.02729},
 primaryClass = {astro-ph.GA},
       adsurl = {https://ui.adsabs.harvard.edu/abs/2025ApJ...985..119Z}
}

@ARTICLE{Zhang_2025c,
       author = {{Zhang}, Zijian and {Li}, Mingyu and {Oguri}, Masamune and {Lin}, Xiaojing and {Inayoshi}, Kohei and {Cerny}, Catherine and {Coe}, Dan and {Diego}, Jose M. and {Fujimoto}, Seiji and {Jiang}, Linhua and {Mahler}, Guillaume and {Matthee}, Jorryt and {Naidu}, Rohan P. and {Sharon}, Keren and {Shen}, Yue and {Zitrin}, Adi and {Abdurro'uf} and {Akins}, Hollis and {Allingham}, Joseph F.~V. and {Amor{\'\i}n}, Ricardo and {Asada}, Yoshihisa and {Atek}, Hakim and {Bauer}, Franz E. and {Brada{\v{c}}}, Maru{\v{s}}a and {Bradley}, Larry D. and {Cai}, Zheng and {Cantalupo}, Sebastiano and {Conselice}, Christopher and {Dai}, Liang and {Dayal}, Pratika and {Egami}, Eiichi and {Eisenstein}, Daniel J. and {Faisst}, Andreas L. and {Fan}, Xiaohui and {Fei}, Qinyue and {Frye}, Brenda L. and {Fudamoto}, Yoshinobu and {Furtak}, Lukas J. and {Golubchik}, Miriam and {Gonz{\'a}lez-Otero}, Mauro and {Harikane}, Yuichi and {Hsiao}, Tiger Yu-Yang and {Jim{\'e}nez-Teja}, Yolanda and {Kartaltepe}, Jeyhan S. and {Kiyota}, Tomokazu and {Koekemoer}, Anton M. and {Kohno}, Kotaro and {Kokorev}, Vasily and {Kumari}, Nimisha and {Labbe}, Ivo and {Lagos}, Claudia D.~P. and {Larison}, Conor and {Liang}, Yongming and {Lucas}, Ray A. and {Lyu}, Jianwei and {Martis}, Nicholas S. and {Magdis}, Georgios E. and {Messa}, Matteo and {Nakane}, Minami and {Noirot}, Ga{\"e}l and {Ortiz}, III, Rafael and {Ouchi}, Masami and {Pierel}, Justin D.~R. and {Postman}, Marc and {Reddy}, Naveen and {Ricotti}, Massimo and {Schaerer}, Daniel and {Schneider}, Raffaella and {Steidel}, Charles C. and {Tee}, Wei Leong and {Tripodi}, Roberta and {Trussler}, James A.~A. and {Umeda}, Hiroya and {Valentino}, Francesco and {Vanzella}, Eros and {Wang}, Feige and {Windhorst}, Rogier and {Wu}, Yunjing and {Wu}, Zihao and {Yanagisawa}, Hiroto and {Yang}, Jinyi and {Sun}, Fengwu},
        title = "{Little red dot variability over a century reveals black hole envelope via a giant Einstein cross}",
      journal = {arXiv e-prints},
         year = 2025,
        month = dec,
          eid = {arXiv:2512.05180},
        pages = {arXiv:2512.05180},
          doi = {10.48550/arXiv.2512.05180},
archivePrefix = {arXiv},
       eprint = {2512.05180},
 primaryClass = {astro-ph.GA},
       adsurl = {https://ui.adsabs.harvard.edu/abs/2025arXiv251205180Z}
}

@ARTICLE{Mazzolari_2024,
       author = {{Mazzolari}, G. and {Gilli}, R. and {Maiolino}, R. and {Prandoni}, I. and {Delvecchio}, I. and {Norman}, C. and {Jimenez-Andrade}, E.~F. and {Belladitta}, S. and {Vito}, F. and {Momjian}, E. and {Chiaberge}, M. and {Trefoloni}, B. and {Signorini}, M. and {Ji}, X. and {D'Amato}, Q. and {Risaliti}, G. and {Baldi}, R.~D. and {Fabian}, A. and {{\"U}bler}, H. and {D'Eugenio}, F. and {Scholtz}, J. and {Juod{\v{z}}balis}, I. and {Mignoli}, M. and {Brusa}, M. and {Murphy}, E. and {Muxlow}, T.~W.~B.},
        title = "{The radio properties of the JWST-discovered AGN}",
      journal = {arXiv e-prints},
         year = 2024,
        month = dec,
          eid = {arXiv:2412.04224},
        pages = {arXiv:2412.04224},
          doi = {10.48550/arXiv.2412.04224},
archivePrefix = {arXiv},
       eprint = {2412.04224},
 primaryClass = {astro-ph.GA},
       adsurl = {https://ui.adsabs.harvard.edu/abs/2024arXiv241204224M}
}

@ARTICLE{Foreman-Mackey_2013,
       author = {{Foreman-Mackey}, Daniel and {Hogg}, David W. and {Lang}, Dustin and {Goodman}, Jonathan},
        title = "{emcee: The MCMC Hammer}",
      journal = {Publications of the Astronomical Society of the Pacific},
         year = 2013,
        month = mar,
       volume = {125},
       number = {925},
        pages = {306},
          doi = {10.1086/670067},
archivePrefix = {arXiv},
       eprint = {1202.3665},
 primaryClass = {astro-ph.IM},
       adsurl = {https://ui.adsabs.harvard.edu/abs/2013PASP..125..306F}
}

@ARTICLE{deGraaff_2025a,
       author = {{de Graaff}, Anna and {Setton}, David J. and {Brammer}, Gabriel and {Cutler}, Sam and {Suess}, Katherine A. and {Labb{\'e}}, Ivo and {Leja}, Joel and {Weibel}, Andrea and {Maseda}, Michael V. and {Whitaker}, Katherine E. and {Bezanson}, Rachel and {Boogaard}, Leindert A. and {Cleri}, Nikko J. and {De Lucia}, Gabriella and {Franx}, Marijn and {Greene}, Jenny E. and {Hirschmann}, Michaela and {Matthee}, Jorryt and {McConachie}, Ian and {Naidu}, Rohan P. and {Oesch}, Pascal A. and {Price}, Sedona H. and {Rix}, Hans-Walter and {Valentino}, Francesco and {Wang}, Bingjie and {Williams}, Christina C.},
        title = "{Efficient formation of a massive quiescent galaxy at redshift 4.9}",
      journal = {Nature Astronomy},
         year = 2025,
        month = feb,
       volume = {9},
        pages = {280-292},
          doi = {10.1038/s41550-024-02424-3},
archivePrefix = {arXiv},
       eprint = {2404.05683},
 primaryClass = {astro-ph.GA},
       adsurl = {https://ui.adsabs.harvard.edu/abs/2025NatAs...9..280D}
}

@ARTICLE{Williams_2024,
       author = {{Williams}, Christina C. and {Alberts}, Stacey and {Ji}, Zhiyuan and {Hainline}, Kevin N. and {Lyu}, Jianwei and {Rieke}, George and {Endsley}, Ryan and {Suess}, Katherine A. and {Sun}, Fengwu and {Johnson}, Benjamin D. and {Florian}, Michael and {Shivaei}, Irene and {Rujopakarn}, Wiphu and {Baker}, William M. and {Bhatawdekar}, Rachana and {Boyett}, Kristan and {Bunker}, Andrew J. and {Cameron}, Alex J. and {Carniani}, Stefano and {Charlot}, Stephane and {Curtis-Lake}, Emma and {DeCoursey}, Christa and {de Graaff}, Anna and {Egami}, Eiichi and {Eisenstein}, Daniel J. and {Gibson}, Justus L. and {Hausen}, Ryan and {Helton}, Jakob M. and {Maiolino}, Roberto and {Maseda}, Michael V. and {Nelson}, Erica J. and {P{\'e}rez-Gonz{\'a}lez}, Pablo G. and {Rieke}, Marcia J. and {Robertson}, Brant E. and {Saxena}, Aayush and {Tacchella}, Sandro and {Willmer}, Christopher N.~A. and {Willott}, Chris J.},
        title = "{The Galaxies Missed by Hubble and ALMA: The Contribution of Extremely Red Galaxies to the Cosmic Census at 3 < z < 8}",
      journal = {The Astrophysical Journal},
         year = 2024,
        month = jun,
       volume = {968},
       number = {1},
          eid = {34},
        pages = {34},
          doi = {10.3847/1538-4357/ad3f17},
archivePrefix = {arXiv},
       eprint = {2311.07483},
 primaryClass = {astro-ph.GA},
       adsurl = {https://ui.adsabs.harvard.edu/abs/2024ApJ...968...34W}
}

@ARTICLE{Furtak_2023a,
       author = {{Furtak}, Lukas J. and {Zitrin}, Adi and {Plat}, Ad{\`e}le and {Fujimoto}, Seiji and {Wang}, Bingjie and {Nelson}, Erica J. and {Labb{\'e}}, Ivo and {Bezanson}, Rachel and {Brammer}, Gabriel B. and {van Dokkum}, Pieter and {Endsley}, Ryan and {Glazebrook}, Karl and {Greene}, Jenny E. and {Leja}, Joel and {Price}, Sedona H. and {Smit}, Renske and {Stark}, Daniel P. and {Weaver}, John R. and {Whitaker}, Katherine E. and {Atek}, Hakim and {Chevallard}, Jacopo and {Curtis-Lake}, Emma and {Dayal}, Pratika and {Feltre}, Anna and {Franx}, Marijn and {Fudamoto}, Yoshinobu and {Marchesini}, Danilo and {Mowla}, Lamiya A. and {Pan}, Richard and {Suess}, Katherine A. and {Vidal-Garc{\'\i}a}, Alba and {Williams}, Christina C.},
        title = "{JWST UNCOVER: Extremely Red and Compact Object at z $_{phot}\sim$  7.6 Triply Imaged by A2744}",
      journal = {The Astrophysical Journal},
         year = 2023,
        month = aug,
       volume = {952},
       number = {2},
          eid = {142},
        pages = {142},
          doi = {10.3847/1538-4357/acdc9d},
archivePrefix = {arXiv},
       eprint = {2212.10531},
 primaryClass = {astro-ph.GA},
       adsurl = {https://ui.adsabs.harvard.edu/abs/2023ApJ...952..142F}
}

@ARTICLE{Greene_2024,
       author = {{Greene}, Jenny E. and {Labbe}, Ivo and {Goulding}, Andy D. and {Furtak}, Lukas J. and {Chemerynska}, Iryna and {Kokorev}, Vasily and {Dayal}, Pratika and {Volonteri}, Marta and {Williams}, Christina C. and {Wang}, Bingjie and {Setton}, David J. and {Burgasser}, Adam J. and {Bezanson}, Rachel and {Atek}, Hakim and {Brammer}, Gabriel and {Cutler}, Sam E. and {Feldmann}, Robert and {Fujimoto}, Seiji and {Glazebrook}, Karl and {de Graaff}, Anna and {Khullar}, Gourav and {Leja}, Joel and {Marchesini}, Danilo and {Maseda}, Michael V. and {Matthee}, Jorryt and {Miller}, Tim B. and {Naidu}, Rohan P. and {Nanayakkara}, Themiya and {Oesch}, Pascal A. and {Pan}, Richard and {Papovich}, Casey and {Price}, Sedona H. and {van Dokkum}, Pieter and {Weaver}, John R. and {Whitaker}, Katherine E. and {Zitrin}, Adi},
        title = "{UNCOVER Spectroscopy Confirms the Surprising Ubiquity of Active Galactic Nuclei in Red Sources at z > 5}",
      journal = {The Astrophysical Journal},
         year = 2024,
        month = mar,
       volume = {964},
       number = {1},
          eid = {39},
        pages = {39},
          doi = {10.3847/1538-4357/ad1e5f},
archivePrefix = {arXiv},
       eprint = {2309.05714},
 primaryClass = {astro-ph.GA},
       adsurl = {https://ui.adsabs.harvard.edu/abs/2024ApJ...964...39G}
}

@ARTICLE{Labbe_2025,
       author = {{Labbe}, Ivo and {Greene}, Jenny E. and {Bezanson}, Rachel and {Fujimoto}, Seiji and {Furtak}, Lukas J. and {Goulding}, Andy D. and {Matthee}, Jorryt and {Naidu}, Rohan P. and {Oesch}, Pascal A. and {Atek}, Hakim and {Brammer}, Gabriel and {Chemerynska}, Iryna and {Coe}, Dan and {Cutler}, Sam E. and {Dayal}, Pratika and {Feldmann}, Robert and {Franx}, Marijn and {Glazebrook}, Karl and {Leja}, Joel and {Maseda}, Michael and {Marchesini}, Danilo and {Nanayakkara}, Themiya and {Nelson}, Erica J. and {Pan}, Richard and {Papovich}, Casey and {Price}, Sedona H. and {Suess}, Katherine A. and {Wang}, Bingjie and {Weaver}, John R. and {Whitaker}, Katherine E. and {Williams}, Christina C. and {Zitrin}, Adi},
        title = "{UNCOVER: Candidate Red Active Galactic Nuclei at 3 < z < 7 with JWST and ALMA}",
      journal = {The Astrophysical Journal},
         year = 2025,
        month = jan,
       volume = {978},
       number = {1},
          eid = {92},
        pages = {92},
          doi = {10.3847/1538-4357/ad3551},
archivePrefix = {arXiv},
       eprint = {2306.07320},
 primaryClass = {astro-ph.GA},
       adsurl = {https://ui.adsabs.harvard.edu/abs/2025ApJ...978...92L}
}

@ARTICLE{Maiolino_2025,
       author = {{Maiolino}, Roberto and {Risaliti}, Guido and {Signorini}, Matilde and {Trefoloni}, Bartolomeo and {Juod{\v{z}}balis}, Ignas and {Scholtz}, Jan and {{\"U}bler}, Hannah and {D'Eugenio}, Francesco and {Carniani}, Stefano and {Fabian}, Andy and {Ji}, Xihan and {Mazzolari}, Giovanni and {Bertola}, Elena and {Brusa}, Marcella and {Bunker}, Andrew J. and {Charlot}, Stephane and {Comastri}, Andrea and {Cresci}, Giovanni and {DeCoursey}, Christa Noel and {Egami}, Eiichi and {Fiore}, Fabrizio and {Gilli}, Roberto and {Perna}, Michele and {Tacchella}, Sandro and {Venturi}, Giacomo},
        title = "{JWST meets Chandra: a large population of Compton thick, feedback-free, and intrinsically X-ray weak AGN, with a sprinkle of SNe}",
      journal = {Monthly Notices of the Royal Astronomical Society},
         year = 2025,
        month = apr,
       volume = {538},
       number = {3},
        pages = {1921-1943},
          doi = {10.1093/mnras/staf359},
archivePrefix = {arXiv},
       eprint = {2405.00504},
 primaryClass = {astro-ph.GA},
       adsurl = {https://ui.adsabs.harvard.edu/abs/2025MNRAS.538.1921M}
}

@ARTICLE{Matthee_2024,
       author = {{Matthee}, Jorryt and {Naidu}, Rohan P. and {Brammer}, Gabriel and {Chisholm}, John and {Eilers}, Anna-Christina and {Goulding}, Andy and {Greene}, Jenny and {Kashino}, Daichi and {Labbe}, Ivo and {Lilly}, Simon J. and {Mackenzie}, Ruari and {Oesch}, Pascal A. and {Weibel}, Andrea and {Wuyts}, Stijn and {Xiao}, Mengyuan and {Bordoloi}, Rongmon and {Bouwens}, Rychard and {van Dokkum}, Pieter and {Illingworth}, Garth and {Kramarenko}, Ivan and {Maseda}, Michael V. and {Mason}, Charlotte and {Meyer}, Romain A. and {Nelson}, Erica J. and {Reddy}, Naveen A. and {Shivaei}, Irene and {Simcoe}, Robert A. and {Yue}, Minghao},
        title = "{Little Red Dots: An Abundant Population of Faint Active Galactic Nuclei at z {\ensuremath{\sim}} 5 Revealed by the EIGER and FRESCO JWST Surveys}",
      journal = {The Astrophysical Journal},
         year = 2024,
        month = mar,
       volume = {963},
       number = {2},
          eid = {129},
        pages = {129},
          doi = {10.3847/1538-4357/ad2345},
archivePrefix = {arXiv},
       eprint = {2306.05448},
 primaryClass = {astro-ph.GA},
       adsurl = {https://ui.adsabs.harvard.edu/abs/2024ApJ...963..129M}
}

@ARTICLE{Kido_2025,
       author = {{Kido}, Daisaburo and {Ioka}, Kunihito and {Hotokezaka}, Kenta and {Inayoshi}, Kohei and {Irwin}, Christopher M.},
        title = "{Black Hole Envelopes in Little Red Dots}",
      journal = {arXiv e-prints},
         year = 2025,
        month = may,
          eid = {arXiv:2505.06965},
        pages = {arXiv:2505.06965},
          doi = {10.48550/arXiv.2505.06965},
archivePrefix = {arXiv},
       eprint = {2505.06965},
 primaryClass = {astro-ph.HE},
       adsurl = {https://ui.adsabs.harvard.edu/abs/2025arXiv250506965K}
}

@ARTICLE{Ji_2025,
       author = {{Ji}, Xihan and {Maiolino}, Roberto and {{\"U}bler}, Hannah and {Scholtz}, Jan and {D'Eugenio}, Francesco and {Sun}, Fengwu and {Perna}, Michele and {Turner}, Hannah and {Arribas}, Santiago and {Bennett}, Jake S. and {Bunker}, Andrew and {Carniani}, Stefano and {Charlot}, St{\'e}phane and {Cresci}, Giovanni and {Curti}, Mirko and {Egami}, Eiichi and {Fabian}, Andy and {Inayoshi}, Kohei and {Isobe}, Yuki and {Jones}, Gareth and {Juod{\v{z}}balis}, Ignas and {Kumari}, Nimisha and {Lyu}, Jianwei and {Mazzolari}, Giovanni and {Parlanti}, Eleonora and {Robertson}, Brant and {Rodr{\'\i}guez Del Pino}, Bruno and {Schneider}, Raffaella and {Sijacki}, Debora and {Tacchella}, Sandro and {Trinca}, Alessandro and {Valiante}, Rosa and {Venturi}, Giacomo and {Volonteri}, Marta and {Willott}, Chris and {Witten}, Callum and {Witstok}, Joris},
        title = "{BlackTHUNDER -- A non-stellar Balmer break in a black hole-dominated little red dot at $z=7.04$}",
      journal = {arXiv e-prints},
         year = 2025,
        month = jan,
          eid = {arXiv:2501.13082},
        pages = {arXiv:2501.13082},
          doi = {10.48550/arXiv.2501.13082},
archivePrefix = {arXiv},
       eprint = {2501.13082},
 primaryClass = {astro-ph.GA},
       adsurl = {https://ui.adsabs.harvard.edu/abs/2025arXiv250113082J}
}

@ARTICLE{Macri_2006,
       author = {{Macri}, L.~M. and {Stanek}, K.~Z. and {Bersier}, D. and {Greenhill}, L.~J. and {Reid}, M.~J.},
        title = "{A New Cepheid Distance to the Maser-Host Galaxy NGC 4258 and Its Implications for the Hubble Constant}",
      journal = {The Astrophysical Journal},
         year = 2006,
        month = dec,
       volume = {652},
       number = {2},
        pages = {1133-1149},
          doi = {10.1086/508530},
archivePrefix = {arXiv},
       eprint = {astro-ph/0608211},
 primaryClass = {astro-ph},
       adsurl = {https://ui.adsabs.harvard.edu/abs/2006ApJ...652.1133M}
}

@ARTICLE{Liu_2026_twinkle,
       author = {{Liu}, Zhaoran and {Naidu}, Rohan P. and {Secunda}, Amy and {Greene}, Jenny E. and {Matthee}, Jorryt and {Chisholm}, John and {de Graaff}, Anna and {Robbins}, Luke and {Antwi-Danso}, Jacqueline and {Brammer}, Gabriel and {Sun}, Wendy Q. and {Eilers}, Anna-Christina and {Fujimoto}, Seiji and {Furtak}, Lukas J. and {Kara}, Erin and {Kokorev}, Vasily and {Marchesini}, Danilo and {Oesch}, Pascal A. and {Pierel}, Justin D.~R. and {Shen}, Xuejian and {Simcoe}, Robert A. and {Torralba}, Alberto and {Vogelsberger}, Mark},
        title = "{How I Wonder What You Are -- JWST's Little Red Dots do not TWINKLE}",
      journal = {arXiv e-prints},
         year = 2026,
        month = apr,
          eid = {arXiv:2604.13000},
        pages = {arXiv:2604.13000},
          doi = {10.48550/arXiv.2604.13000},
archivePrefix = {arXiv},
       eprint = {2604.13000},
 primaryClass = {astro-ph.GA},
       adsurl = {https://ui.adsabs.harvard.edu/abs/2026arXiv260413000L}
}

@ARTICLE{Cantiello_2025,
       author = {{Cantiello}, Matteo and {Hassan}, Jake B. and {Perna}, Rosalba and {Armitage}, Philip J. and {Begelman}, Mitchell C. and {Jiang}, Yan-Fei and {Ryu}, Taeho and {Townsend}, Richard H.~D.},
        title = "{Pulsational Instability of Quasi-Stars: Interpreting the Variability of Little Red Dots}",
      journal = {arXiv e-prints},
         year = 2025,
        month = dec,
          eid = {arXiv:2512.17997},
        pages = {arXiv:2512.17997},
          doi = {10.48550/arXiv.2512.17997},
archivePrefix = {arXiv},
       eprint = {2512.17997},
 primaryClass = {astro-ph.HE},
       adsurl = {https://ui.adsabs.harvard.edu/abs/2025arXiv251217997C}
}

@ARTICLE{Baggen_2025,
       author = {{Baggen}, Josephine F.~W. and {van Dokkum}, Pieter and {Labb{\'e}}, Ivo and {Brammer}, Gabriel},
        title = "{(Re)solving the complex multi-scale morphology and V-shaped SED of a newly discovered strongly-lensed Little Red Dot in Abell 383}",
      journal = {arXiv e-prints},
         year = 2025,
        month = dec,
          eid = {arXiv:2512.03239},
        pages = {arXiv:2512.03239},
          doi = {10.48550/arXiv.2512.03239},
archivePrefix = {arXiv},
       eprint = {2512.03239},
 primaryClass = {astro-ph.GA},
       adsurl = {https://ui.adsabs.harvard.edu/abs/2025arXiv251203239B}
}

@ARTICLE{Tee_2025,
       author = {{Tee}, Wei Leong and {Fan}, Xiaohui and {Wang}, Feige and {Yang}, Jinyi},
        title = "{Lack of Rest-frame Ultraviolet Variability in Little Red Dots Based on HST and JWST Observations}",
      journal = {The Astrophysical Journal Letter},
         year = 2025,
        month = apr,
       volume = {983},
       number = {1},
          eid = {L26},
        pages = {L26},
          doi = {10.3847/2041-8213/adc5e3},
archivePrefix = {arXiv},
       eprint = {2412.05242},
 primaryClass = {astro-ph.GA},
       adsurl = {https://ui.adsabs.harvard.edu/abs/2025ApJ...983L..26T}
}

@ARTICLE{Riess_2022_SHOES,
       author = {{Riess}, Adam G. and {Yuan}, Wenlong and {Macri}, Lucas M. and {Scolnic}, Dan and {Brout}, Dillon and {Casertano}, Stefano and {Jones}, David O. and {Murakami}, Yukei and {Anand}, Gagandeep S. and {Breuval}, Louise and {Brink}, Thomas G. and {Filippenko}, Alexei V. and {Hoffmann}, Samantha and {Jha}, Saurabh W. and {D'arcy Kenworthy}, W. and {Mackenty}, John and {Stahl}, Benjamin E. and {Zheng}, WeiKang},
        title = "{A Comprehensive Measurement of the Local Value of the Hubble Constant with 1 km s$^{-1}$ Mpc$^{-1}$ Uncertainty from the Hubble Space Telescope and the SH0ES Team}",
      journal = {The Astrophysical Journal Letter},
         year = 2022,
        month = jul,
       volume = {934},
       number = {1},
          eid = {L7},
        pages = {L7},
          doi = {10.3847/2041-8213/ac5c5b},
archivePrefix = {arXiv},
       eprint = {2112.04510},
 primaryClass = {astro-ph.CO},
       adsurl = {https://ui.adsabs.harvard.edu/abs/2022ApJ...934L...7R}
}

@ARTICLE{Planck_2020,
       author = {{Planck Collaboration} and {Aghanim}, N. and {Akrami}, Y. and {Ashdown}, M. and {Aumont}, J. and {Baccigalupi}, C. and {Ballardini}, M. and {Banday}, A.~J. and {Barreiro}, R.~B. and {Bartolo}, N. and {Basak}, S. and {Battye}, R. and {Benabed}, K. and {Bernard}, J.-P. and {Bersanelli}, M. and {Bielewicz}, P. and {Bock}, J.~J. and {Bond}, J.~R. and {Borrill}, J. and {Bouchet}, F.~R. and {Boulanger}, F. and {Bucher}, M. and {Burigana}, C. and {Butler}, R.~C. and {Calabrese}, E. and {Cardoso}, J.-F. and {Carron}, J. and {Challinor}, A. and {Chiang}, H.~C. and {Chluba}, J. and {Colombo}, L.~P.~L. and {Combet}, C. and {Contreras}, D. and {Crill}, B.~P. and {Cuttaia}, F. and {de Bernardis}, P. and {de Zotti}, G. and {Delabrouille}, J. and {Delouis}, J.-M. and {Di Valentino}, E. and {Diego}, J.~M. and {Dor{\'e}}, O. and {Douspis}, M. and {Ducout}, A. and {Dupac}, X. and {Dusini}, S. and {Efstathiou}, G. and {Elsner}, F. and {En{\ss}lin}, T.~A. and {Eriksen}, H.~K. and {Fantaye}, Y. and {Farhang}, M. and {Fergusson}, J. and {Fernandez-Cobos}, R. and {Finelli}, F. and {Forastieri}, F. and {Frailis}, M. and {Fraisse}, A.~A. and {Franceschi}, E. and {Frolov}, A. and {Galeotta}, S. and {Galli}, S. and {Ganga}, K. and {G{\'e}nova-Santos}, R.~T. and {Gerbino}, M. and {Ghosh}, T. and {Gonz{\'a}lez-Nuevo}, J. and {G{\'o}rski}, K.~M. and {Gratton}, S. and {Gruppuso}, A. and {Gudmundsson}, J.~E. and {Hamann}, J. and {Handley}, W. and {Hansen}, F.~K. and {Herranz}, D. and {Hildebrandt}, S.~R. and {Hivon}, E. and {Huang}, Z. and {Jaffe}, A.~H. and {Jones}, W.~C. and {Karakci}, A. and {Keih{\"a}nen}, E. and {Keskitalo}, R. and {Kiiveri}, K. and {Kim}, J. and {Kisner}, T.~S. and {Knox}, L. and {Krachmalnicoff}, N. and {Kunz}, M. and {Kurki-Suonio}, H. and {Lagache}, G. and {Lamarre}, J.-M. and {Lasenby}, A. and {Lattanzi}, M. and {Lawrence}, C.~R. and {Le Jeune}, M. and {Lemos}, P. and {Lesgourgues}, J. and {Levrier}, F. and {Lewis}, A. and {Liguori}, M. and {Lilje}, P.~B. and {Lilley}, M. and {Lindholm}, V. and {L{\'o}pez-Caniego}, M. and {Lubin}, P.~M. and {Ma}, Y.-Z. and {Mac{\'\i}as-P{\'e}rez}, J.~F. and {Maggio}, G. and {Maino}, D. and {Mandolesi}, N. and {Mangilli}, A. and {Marcos-Caballero}, A. and {Maris}, M. and {Martin}, P.~G. and {Martinelli}, M. and {Mart{\'\i}nez-Gonz{\'a}lez}, E. and {Matarrese}, S. and {Mauri}, N. and {McEwen}, J.~D. and {Meinhold}, P.~R. and {Melchiorri}, A. and {Mennella}, A. and {Migliaccio}, M. and {Millea}, M. and {Mitra}, S. and {Miville-Desch{\^e}nes}, M.-A. and {Molinari}, D. and {Montier}, L. and {Morgante}, G. and {Moss}, A. and {Natoli}, P. and {N{\o}rgaard-Nielsen}, H.~U. and {Pagano}, L. and {Paoletti}, D. and {Partridge}, B. and {Patanchon}, G. and {Peiris}, H.~V. and {Perrotta}, F. and {Pettorino}, V. and {Piacentini}, F. and {Polastri}, L. and {Polenta}, G. and {Puget}, J.-L. and {Rachen}, J.~P. and {Reinecke}, M. and {Remazeilles}, M. and {Renzi}, A. and {Rocha}, G. and {Rosset}, C. and {Roudier}, G. and {Rubi{\~n}o-Mart{\'\i}n}, J.~A. and {Ruiz-Granados}, B. and {Salvati}, L. and {Sandri}, M. and {Savelainen}, M. and {Scott}, D. and {Shellard}, E.~P.~S. and {Sirignano}, C. and {Sirri}, G. and {Spencer}, L.~D. and {Sunyaev}, R. and {Suur-Uski}, A.-S. and {Tauber}, J.~A. and {Tavagnacco}, D. and {Tenti}, M. and {Toffolatti}, L. and {Tomasi}, M. and {Trombetti}, T. and {Valenziano}, L. and {Valiviita}, J. and {Van Tent}, B. and {Vibert}, L. and {Vielva}, P. and {Villa}, F. and {Vittorio}, N. and {Wandelt}, B.~D. and {Wehus}, I.~K. and {White}, M. and {White}, S.~D.~M. and {Zacchei}, A. and {Zonca}, A.},
        title = "{Planck 2018 results. VI. Cosmological parameters}",
      journal = {A\&A},
         year = 2020,
        month = sep,
       volume = {641},
          eid = {A6},
        pages = {A6},
          doi = {10.1051/0004-6361/201833910},
archivePrefix = {arXiv},
       eprint = {1807.06209},
 primaryClass = {astro-ph.CO},
       adsurl = {https://ui.adsabs.harvard.edu/abs/2020A&A...641A...6P}
}

@ARTICLE{Freedman_2020_TRGB,
       author = {{Freedman}, Wendy L. and {Madore}, Barry F. and {Hoyt}, Taylor and {Jang}, In Sung and {Beaton}, Rachael and {Lee}, Myung Gyoon and {Monson}, Andrew and {Neeley}, Jill and {Rich}, Jeffrey},
        title = "{Calibration of the Tip of the Red Giant Branch}",
      journal = {The Astrophysical Journal},
         year = 2020,
        month = mar,
       volume = {891},
       number = {1},
          eid = {57},
        pages = {57},
          doi = {10.3847/1538-4357/ab7339},
archivePrefix = {arXiv},
       eprint = {2002.01550},
 primaryClass = {astro-ph.GA},
       adsurl = {https://ui.adsabs.harvard.edu/abs/2020ApJ...891...57F}
}

@ARTICLE{Wong_2020_lensingH0,
       author = {{Wong}, Kenneth C. and {Suyu}, Sherry H. and {Chen}, Geoff C.-F. and {Rusu}, Cristian E. and {Millon}, Martin and {Sluse}, Dominique and {Bonvin}, Vivien and {Fassnacht}, Christopher D. and {Taubenberger}, Stefan and {Auger}, Matthew W. and {Birrer}, Simon and {Chan}, James H.~H. and {Courbin}, Frederic and {Hilbert}, Stefan and {Tihhonova}, Olga and {Treu}, Tommaso and {Agnello}, Adriano and {Ding}, Xuheng and {Jee}, Inh and {Komatsu}, Eiichiro and {Shajib}, Anowar J. and {Sonnenfeld}, Alessandro and {Blandford}, Roger D. and {Koopmans}, L{\'e}on V.~E. and {Marshall}, Philip J. and {Meylan}, Georges},
        title = "{H0LiCOW - XIII. A 2.4 per cent measurement of H$_{0}$ from lensed quasars: 5.3{\ensuremath{\sigma}} tension between early- and late-Universe probes}",
      journal = {Monthly Notices of the Royal Astronomical Society},
         year = 2020,
        month = oct,
       volume = {498},
       number = {1},
        pages = {1420-1439},
          doi = {10.1093/mnras/stz3094},
archivePrefix = {arXiv},
       eprint = {1907.04869},
 primaryClass = {astro-ph.CO},
       adsurl = {https://ui.adsabs.harvard.edu/abs/2020MNRAS.498.1420W}
}

@ARTICLE{Ji_2026,
       author = {{Ji}, Xihan and {D'Eugenio}, Francesco and {Juod{\v{z}}balis}, Ignas and {Walton}, Dominic J. and {Fabian}, Andrew C. and {Maiolino}, Roberto and {Ramos Almeida}, Cristina and {Acosta Pulido}, Jose A. and {Belokurov}, Vasily A. and {Isobe}, Yuki and {Jones}, Gareth and {Maraston}, Claudia and {Scholtz}, Jan and {Simmonds}, Charlotte and {Tacchella}, Sandro and {Terlevich}, Elena and {Terlevich}, Roberto},
        title = "{Lord of LRDs: insights into a 'Little Red Dot' with a low-ionization spectrum at z = 0.1}",
      journal = {Monthly Notices of the Royal Astronomical Society},
         year = 2026,
        month = jan,
       volume = {545},
       number = {3},
          eid = {staf2235},
        pages = {staf2235},
          doi = {10.1093/mnras/staf2235},
archivePrefix = {arXiv},
       eprint = {2507.23774},
 primaryClass = {astro-ph.GA},
       adsurl = {https://ui.adsabs.harvard.edu/abs/2026MNRAS.545S2235J}
}

@ARTICLE{Juodzbalis_2024,
       author = {{Juod{\v{z}}balis}, Ignas and {Ji}, Xihan and {Maiolino}, Roberto and {D'Eugenio}, Francesco and {Scholtz}, Jan and {Risaliti}, Guido and {Fabian}, Andrew C. and {Mazzolari}, Giovanni and {Gilli}, Roberto and {Prandoni}, Isabella and {Arribas}, Santiago and {Bunker}, Andrew J. and {Carniani}, Stefano and {Charlot}, St{\'e}phane and {Curtis-Lake}, Emma and {de Graaff}, Anna and {Hainline}, Kevin and {Parlanti}, Eleonora and {Perna}, Michele and {P{\'e}rez-Gonz{\'a}lez}, Pablo G. and {Robertson}, Brant and {Tacchella}, Sandro and {{\"U}bler}, Hannah and {Williams}, Christina C. and {Willott}, Chris and {Witstok}, Joris},
        title = "{JADES - the Rosetta stone of JWST-discovered AGN: deciphering the intriguing nature of early AGN}",
      journal = {Monthly Notices of the Royal Astronomical Society},
         year = 2024,
        month = nov,
       volume = {535},
       number = {1},
        pages = {853-873},
          doi = {10.1093/mnras/stae2367},
archivePrefix = {arXiv},
       eprint = {2407.08643},
 primaryClass = {astro-ph.GA},
       adsurl = {https://ui.adsabs.harvard.edu/abs/2024MNRAS.535..853J}
}

@ARTICLE{Leavitt_1912,
       author = {{Leavitt}, Henrietta S. and {Pickering}, Edward C.},
        title = "{Periods of 25 Variable Stars in the Small Magellanic Cloud.}",
      journal = {Harvard College Observatory Circular},
         year = 1912,
        month = mar,
       volume = {173},
        pages = {1-3},
       adsurl = {https://ui.adsabs.harvard.edu/abs/1912HarCi.173....1L}
}

@ARTICLE{Cenarro2002,
       author = {{Cenarro}, A.~J. and {Gorgas}, J. and {Cardiel}, N. and {Vazdekis}, A. and {Peletier}, R.~F.},
        title = "{Empirical calibration of the near-infrared Ca II triplet - III. Fitting functions}",
      journal = {Monthly Notices of the Royal Astronomical Society},
         year = 2002,
        month = feb,
       volume = {329},
       number = {4},
        pages = {863-876},
          doi = {10.1046/j.1365-8711.2002.05029.x},
archivePrefix = {arXiv},
       eprint = {astro-ph/0112448},
 primaryClass = {astro-ph},
       adsurl = {https://ui.adsabs.harvard.edu/abs/2002MNRAS.329..863C}
}

@ARTICLE{DiValentino_2021,
       author = {{Di Valentino}, Eleonora and {Mena}, Olga and {Pan}, Supriya and {Visinelli}, Luca and {Yang}, Weiqiang and {Melchiorri}, Alessandro and {Mota}, David F. and {Riess}, Adam G. and {Silk}, Joseph},
        title = "{In the realm of the Hubble tension-a review of solutions}",
      journal = {Classical and Quantum Gravity},
         year = 2021,
        month = jul,
       volume = {38},
       number = {15},
          eid = {153001},
        pages = {153001},
          doi = {10.1088/1361-6382/ac086d},
archivePrefix = {arXiv},
       eprint = {2103.01183},
 primaryClass = {astro-ph.CO},
       adsurl = {https://ui.adsabs.harvard.edu/abs/2021CQGra..38o3001D}
}

@ARTICLE{Lin_2024,
       author = {{Lin}, Xiaojing and {Wang}, Feige and {Fan}, Xiaohui and {Cai}, Zheng and {Champagne}, Jaclyn B. and {Sun}, Fengwu and {Volonteri}, Marta and {Yang}, Jinyi and {Hennawi}, Joseph F. and {Ba{\~n}ados}, Eduardo and {Barth}, Aaron and {Eilers}, Anna-Christina and {Farina}, Emanuele Paolo and {Liu}, Weizhe and {Jin}, Xiangyu and {Jun}, Hyunsung D. and {Lupi}, Alessandro and {Kakiichi}, Koki and {Mazzucchelli}, Chiara and {Onoue}, Masafusa and {Pan}, Zhiwei and {Pizzati}, Elia and {Rojas-Ruiz}, Sof{\'\i}a and {Schindler}, Jan-Torge and {Trakhtenbrot}, Benny and {Shen}, Yue and {Trebitsch}, Maxime and {Zhuang}, Ming-Yang and {Endsley}, Ryan and {Meyer}, Romain A. and {Li}, Zihao and {Li}, Mingyu and {Pudoka}, Maria and {Tee}, Wei Leong and {Wu}, Yunjing and {Zhang}, Haowen},
        title = "{A SPectroscopic Survey of Biased Halos In the Reionization Era (ASPIRE): Broad-line AGN at z = 4‑5 Revealed by JWST/NIRCam WFSS}",
      journal = {The Astrophysical Journal},
         year = 2024,
        month = oct,
       volume = {974},
       number = {1},
          eid = {147},
        pages = {147},
          doi = {10.3847/1538-4357/ad6565},
archivePrefix = {arXiv},
       eprint = {2407.17570},
 primaryClass = {astro-ph.GA},
       adsurl = {https://ui.adsabs.harvard.edu/abs/2024ApJ...974..147L}
}

@ARTICLE{Ananna_2024,
       author = {{Ananna}, Tonima Tasnim and {Bogd{\'a}n}, {\'A}kos and {Kov{\'a}cs}, Orsolya E. and {Natarajan}, Priyamvada and {Hickox}, Ryan C.},
        title = "{X-Ray View of Little Red Dots: Do They Host Supermassive Black Holes?}",
      journal = {The Astrophysical Journal Letter},
         year = 2024,
        month = jul,
       volume = {969},
       number = {1},
          eid = {L18},
        pages = {L18},
          doi = {10.3847/2041-8213/ad5669},
archivePrefix = {arXiv},
       eprint = {2404.19010},
 primaryClass = {astro-ph.GA},
       adsurl = {https://ui.adsabs.harvard.edu/abs/2024ApJ...969L..18A}
}

@ARTICLE{Yue_2024,
       author = {{Yue}, Minghao and {Eilers}, Anna-Christina and {Ananna}, Tonima Tasnim and {Panagiotou}, Christos and {Kara}, Erin and {Miyaji}, Takamitsu},
        title = "{Stacking X-Ray Observations of ``Little Red Dots'': Implications for Their Active Galactic Nucleus Properties}",
      journal = {The Astrophysical Journal Letter},
         year = 2024,
        month = oct,
       volume = {974},
       number = {2},
          eid = {L26},
        pages = {L26},
          doi = {10.3847/2041-8213/ad7eba},
archivePrefix = {arXiv},
       eprint = {2404.13290},
 primaryClass = {astro-ph.GA},
       adsurl = {https://ui.adsabs.harvard.edu/abs/2024ApJ...974L..26Y}
}

@ARTICLE{Suyu_2013,
       author = {{Suyu}, S.~H. and {Auger}, M.~W. and {Hilbert}, S. and {Marshall}, P.~J. and {Tewes}, M. and {Treu}, T. and {Fassnacht}, C.~D. and {Koopmans}, L.~V.~E. and {Sluse}, D. and {Blandford}, R.~D. and {Courbin}, F. and {Meylan}, G.},
        title = "{Two Accurate Time-delay Distances from Strong Lensing: Implications for Cosmology}",
      journal = {The Astrophysical Journal},
         year = 2013,
        month = apr,
       volume = {766},
       number = {2},
          eid = {70},
        pages = {70},
          doi = {10.1088/0004-637X/766/2/70},
archivePrefix = {arXiv},
       eprint = {1208.6010},
 primaryClass = {astro-ph.CO},
       adsurl = {https://ui.adsabs.harvard.edu/abs/2013ApJ...766...70S}
}

@ARTICLE{Greene_2013,
       author = {{Greene}, Zach S. and {Suyu}, Sherry H. and {Treu}, Tommaso and {Hilbert}, Stefan and {Auger}, Matthew W. and {Collett}, Thomas E. and {Marshall}, Philip J. and {Fassnacht}, Christopher D. and {Blandford}, Roger D. and {Brada{\v{c}}}, Maru{\v{s}}a and {Koopmans}, L{\'e}on V.~E.},
        title = "{Improving the Precision of Time-delay Cosmography with Observations of Galaxies along the Line of Sight}",
      journal = {The Astrophysical Journal},
         year = 2013,
        month = may,
       volume = {768},
       number = {1},
          eid = {39},
        pages = {39},
          doi = {10.1088/0004-637X/768/1/39},
archivePrefix = {arXiv},
       eprint = {1303.3588},
 primaryClass = {astro-ph.CO},
       adsurl = {https://ui.adsabs.harvard.edu/abs/2013ApJ...768...39G}
}

@ARTICLE{Falco_1985,
       author = {{Falco}, E.~E. and {Gorenstein}, M.~V. and {Shapiro}, I.~I.},
        title = "{On model-dependent bounds on H 0 from gravitational images : application to Q 0957+561 A, B.}",
      journal = {The Astrophysical Journal Letter},
         year = 1985,
        month = feb,
       volume = {289},
        pages = {L1-L4},
          doi = {10.1086/184422},
       adsurl = {https://ui.adsabs.harvard.edu/abs/1985ApJ...289L...1F}
}

@ARTICLE{Hoyle_1963,
       author = {{Hoyle}, F. and {Fowler}, W.~A.},
        title = "{On the nature of strong radio sources}",
      journal = {Monthly Notices of the Royal Astronomical Society},
         year = 1963,
        month = jan,
       volume = {125},
        pages = {169},
          doi = {10.1093/mnras/125.2.169},
       adsurl = {https://ui.adsabs.harvard.edu/abs/1963MNRAS.125..169H}
}

@ARTICLE{Pandey_2025,
       author = {{Pandey}, Rakesh and {Palau}, Aina and {Serna}, Javier and {Kuiper}, Rolf and {S{\'a}nchez-Monge}, Alvaro and {Sharma}, Saurabh and {Sahai}, Raghvendra and {Contreras}, Carmen S{\'a}nchez and {Hern{\'a}ndez}, Jes{\'u}s and {Rom{\'a}n-Z{\'u}{\~n}iga}, Carlos and {Rodler}, Florian},
        title = "{Testing the bloated star hypothesis in the massive young stellar object IRAS 19520+2759 through optical and infrared variability}",
      journal = {Monthly Notices of the Royal Astronomical Society},
         year = 2025,
        month = mar,
       volume = {541},
       number = {4},
        pages = {3772-3786},
          doi = {10.1093/mnras/staf493},
archivePrefix = {arXiv},
       eprint = {2503.16733},
 primaryClass = {astro-ph.SR},
       adsurl = {https://ui.adsabs.harvard.edu/abs/2025MNRAS.541.3772P}
}

@ARTICLE{Inayoshi_2013b,
       author = {{Inayoshi}, Kohei and {Sugiyama}, Koichiro and {Hosokawa}, Takashi and {Motogi}, Kazuhito and {Tanaka}, Kei E.~I.},
        title = "{Direct Diagnostics of Forming Massive Stars: Stellar Pulsation and Periodic Variability of Maser Sources}",
      journal = {The Astrophysical Journal Letter},
         year = 2013,
        month = jun,
       volume = {769},
       number = {2},
          eid = {L20},
        pages = {L20},
          doi = {10.1088/2041-8205/769/2/L20},
archivePrefix = {arXiv},
       eprint = {1304.5241},
 primaryClass = {astro-ph.SR},
       adsurl = {https://ui.adsabs.harvard.edu/abs/2013ApJ...769L..20I}
}

@ARTICLE{Inayoshi_2013,
   author = {{Inayoshi}, K. and {Hosokawa}, T. and {Omukai}, K.},
    title = "{Pulsational instability of supergiant protostars: do they grow supermassive by accretion?}",
  journal = {Monthly Notices of the Royal Astronomical Society},
archivePrefix = "arXiv",
   eprint = {1302.6065},
 primaryClass = "astro-ph.SR",
     year = 2013,
    month = jun,
   volume = 431,
    pages = {3036-3044},
      doi = {10.1093/mnras/stt362},
   adsurl = {http://adsabs.harvard.edu/abs/2013MNRAS.431.3036I}
}

@ARTICLE{Begelman_2006,
   author = {{Begelman}, M.~C. and {Volonteri}, M. and {Rees}, M.~J.},
    title = "{Formation of supermassive black holes by direct collapse in pre-galactic haloes}",
  journal = {Monthly Notices of the Royal Astronomical Society},
   eprint = {astro-ph/0602363},
     year = 2006,
    month = jul,
   volume = 370,
    pages = {289-298},
      doi = {10.1111/j.1365-2966.2006.10467.x},
   adsurl = {http://adsabs.harvard.edu/abs/2006MNRAS.370..289B}
}

@ARTICLE{Schwarzschild_1941,
       author = {{Schwarzschild}, Martin},
        title = "{Overtone Pulsations for the Standard Model.}",
      journal = {The Astrophysical Journal},
         year = 1941,
        month = sep,
       volume = {94},
        pages = {245},
          doi = {10.1086/144329},
       adsurl = {https://ui.adsabs.harvard.edu/abs/1941ApJ....94..245S}
}

@ARTICLE{Bond_1991,
       author = {{Bond}, J.~R. and {Cole}, S. and {Efstathiou}, G. and {Kaiser}, N.},
        title = "{Excursion Set Mass Functions for Hierarchical Gaussian Fluctuations}",
      journal = {The Astrophysical Journal},
         year = 1991,
        month = oct,
       volume = {379},
        pages = {440},
          doi = {10.1086/170520},
       adsurl = {https://ui.adsabs.harvard.edu/abs/1991ApJ...379..440B}
}

@ARTICLE{Perez-Gonzalez_2024,
       author = {{P{\'e}rez-Gonz{\'a}lez}, Pablo G. and {Barro}, Guillermo and {Rieke}, George H. and {Lyu}, Jianwei and {Rieke}, Marcia and {Alberts}, Stacey and {Williams}, Christina C. and {Hainline}, Kevin and {Sun}, Fengwu and {Pusk{\'a}s}, D{\'a}vid and {Annunziatella}, Marianna and {Baker}, William M. and {Bunker}, Andrew J. and {Egami}, Eiichi and {Ji}, Zhiyuan and {Johnson}, Benjamin D. and {Robertson}, Brant and {Rodr{\'\i}guez Del Pino}, Bruno and {Rujopakarn}, Wiphu and {Shivaei}, Irene and {Tacchella}, Sandro and {Willmer}, Christopher N.~A. and {Willott}, Chris},
        title = "{What Is the Nature of Little Red Dots and what Is Not, MIRI SMILES Edition}",
      journal = {The Astrophysical Journal},
         year = 2024,
        month = jun,
       volume = {968},
       number = {1},
          eid = {4},
        pages = {4},
          doi = {10.3847/1538-4357/ad38bb},
archivePrefix = {arXiv},
       eprint = {2401.08782},
 primaryClass = {astro-ph.GA},
       adsurl = {https://ui.adsabs.harvard.edu/abs/2024ApJ...968....4P}
}

@ARTICLE{Hosokawa_2013,
   author = {{Hosokawa}, T. and {Yorke}, H.~W. and {Inayoshi}, K. and {Omukai}, K. and 
	{Yoshida}, N.},
    title = "{Formation of Primordial Supermassive Stars by Rapid Mass Accretion}",
  journal = {The Astrophysical Journal},
archivePrefix = "arXiv",
   eprint = {1308.4457},
 primaryClass = "astro-ph.SR",
     year = 2013,
    month = dec,
   volume = 778,
      eid = {178},
    pages = {178},
      doi = {10.1088/0004-637X/778/2/178},
   adsurl = {http://adsabs.harvard.edu/abs/2013ApJ...778..178H}
}

@ARTICLE{Eddington_1917,
       author = {{Eddington}, A.~S.},
        title = "{The pulsation theory of Cepheid variables}",
      journal = {The Observatory},
         year = 1917,
        month = aug,
       volume = {40},
        pages = {290-293},
       adsurl = {https://ui.adsabs.harvard.edu/abs/1917Obs....40..290E}
}

@BOOK{Cox_1980,
       author = {{Cox}, John P.},
        title = "{Theory of Stellar Pulsation. (PSA-2), Volume 2}",
         year = 1980,
       volume = {2},
       adsurl = {https://ui.adsabs.harvard.edu/abs/1980tsp..book.....C}
}

@ARTICLE{Li_1994,
       author = {{Li}, Y. and {Gong}, Z.~G.},
        title = "{Red supergiant variables in the Large Magellanic Cloud : their evolution and pulsations.}",
      journal = {A\&A},
         year = 1994,
        month = sep,
       volume = {289},
        pages = {449-457},
       adsurl = {https://ui.adsabs.harvard.edu/abs/1994A&A...289..449L}
}

@BOOK{Chandrasekhar_1939,
       author = {{Chandrasekhar}, Subrahmanyan},
        title = "{An introduction to the study of stellar structure}",
         year = 1939,
       adsurl = {https://ui.adsabs.harvard.edu/abs/1939isss.book.....C}
}

@ARTICLE{Barro_2024,
       author = {{Barro}, Guillermo and {P{\'e}rez-Gonz{\'a}lez}, Pablo G. and {Kocevski}, Dale D. and {McGrath}, Elizabeth J. and {Trump}, Jonathan R. and {Simons}, Raymond C. and {Somerville}, Rachel S. and {Yung}, L.~Y. Aaron and {Arrabal Haro}, Pablo and {Akins}, Hollis B. and {Bagley}, Michaela B. and {Cleri}, Nikko J. and {Costantin}, Luca and {Davis}, Kelcey and {Dickinson}, Mark and {Finkelstein}, Steve L. and {Giavalisco}, Mauro and {G{\'o}mez-Guijarro}, Carlos and {Hathi}, Nimish P. and {Hirschmann}, Michaela and {Holwerda}, Benne W. and {Huertas-Company}, Marc and {Kartaltepe}, Jeyhan S. and {Koekemoer}, Anton M. and {Lucas}, Ray A. and {Papovich}, Casey and {Pirzkal}, Nor and {Seill{\'e}}, Lise-Marie and {Tacchella}, Sandro and {Wuyts}, Stijn and {Wilkins}, Stephen M. and {de la Vega}, Alexander and {Yang}, Guang and {Zavala}, Jorge A.},
        title = "{Extremely Red Galaxies at z = 5{\textendash}9 with MIRI and NIRSpec: Dusty Galaxies or Obscured Active Galactic Nuclei?}",
      journal = {The Astrophysical Journal},
         year = 2024,
        month = mar,
       volume = {963},
       number = {2},
          eid = {128},
        pages = {128},
          doi = {10.3847/1538-4357/ad167e},
archivePrefix = {arXiv},
       eprint = {2305.14418},
 primaryClass = {astro-ph.GA},
       adsurl = {https://ui.adsabs.harvard.edu/abs/2024ApJ...963..128B}
}

@ARTICLE{Harikane_2023_agn,
       author = {{Harikane}, Yuichi and {Zhang}, Yechi and {Nakajima}, Kimihiko and {Ouchi}, Masami and {Isobe}, Yuki and {Ono}, Yoshiaki and {Hatano}, Shun and {Xu}, Yi and {Umeda}, Hiroya},
        title = "{A JWST/NIRSpec First Census of Broad-line AGNs at z = 4-7: Detection of 10 Faint AGNs with M $_{BH}$ {}10$^{6}$-{}10$^{8}$ M $_{{\ensuremath{\odot}}}$ and Their Host Galaxy Properties}",
      journal = {The Astrophysical Journal},
         year = 2023,
        month = dec,
       volume = {959},
       number = {1},
          eid = {39},
        pages = {39},
          doi = {10.3847/1538-4357/ad029e},
archivePrefix = {arXiv},
       eprint = {2303.11946},
 primaryClass = {astro-ph.GA},
       adsurl = {https://ui.adsabs.harvard.edu/abs/2023ApJ...959...39H}
}

@ARTICLE{Kocevski_2025,
       author = {{Kocevski}, Dale D. and {Finkelstein}, Steven L. and {Barro}, Guillermo and {Taylor}, Anthony J. and {Calabr{\`o}}, Antonello and {Laloux}, Brivael and {Buchner}, Johannes and {Trump}, Jonathan R. and {Leung}, Gene C.~K. and {Yang}, Guang and {Dickinson}, Mark and {P{\'e}rez-Gonz{\'a}lez}, Pablo G. and {Pacucci}, Fabio and {Inayoshi}, Kohei and {Somerville}, Rachel S. and {McGrath}, Elizabeth J. and {Akins}, Hollis B. and {Bagley}, Micaela B. and {Bowler}, Rebecca A.~A. and {Bisigello}, Laura and {Carnall}, Adam and {Casey}, Caitlin M. and {Cheng}, Yingjie and {Cleri}, Nikko J. and {Costantin}, Luca and {Cullen}, Fergus and {Davis}, Kelcey and {Donnan}, Callum T. and {Dunlop}, James S. and {Ellis}, Richard S. and {Ferguson}, Henry C. and {Fujimoto}, Seiji and {Fontana}, Adriano and {Giavalisco}, Mauro and {Grazian}, Andrea and {Grogin}, Norman A. and {Hathi}, Nimish P. and {Hirschmann}, Michaela and {Huertas-Company}, Marc and {Holwerda}, Benne W. and {Illingworth}, Garth and {Juneau}, St{\'e}phanie and {Kartaltepe}, Jeyhan S. and {Koekemoer}, Anton M. and {Li}, Wenxiu and {Lucas}, Ray A. and {Magee}, Dan and {Mason}, Charlotte and {McLeod}, Derek J. and {McLure}, Ross J. and {Napolitano}, Lorenzo and {Papovich}, Casey and {Pirzkal}, Nor and {Rodighiero}, Giulia and {Santini}, Paola and {Wilkins}, Stephen M. and {Yung}, L.~Y. Aaron},
        title = "{The Rise of Faint, Red Active Galactic Nuclei at z > 4: A Sample of Little Red Dots in the JWST Extragalactic Legacy Fields}",
      journal = {The Astrophysical Journal},
         year = 2025,
        month = jun,
       volume = {986},
       number = {2},
          eid = {126},
        pages = {126},
          doi = {10.3847/1538-4357/adbc7d},
archivePrefix = {arXiv},
       eprint = {2404.03576},
 primaryClass = {astro-ph.GA},
       adsurl = {https://ui.adsabs.harvard.edu/abs/2025ApJ...986..126K}
}

@ARTICLE{Akins_2025,
       author = {{Akins}, Hollis B. and {Casey}, Caitlin M. and {Lambrides}, Erini and {Allen}, Natalie and {Andika}, Irham T. and {Brinch}, Malte and {Champagne}, Jaclyn B. and {Cooper}, Olivia and {Ding}, Xuheng and {Drakos}, Nicole E. and {Faisst}, Andreas and {Finkelstein}, Steven L. and {Franco}, Maximilien and {Fujimoto}, Seiji and {Gentile}, Fabrizio and {Gillman}, Steven and {Gozaliasl}, Ghassem and {Harish}, Santosh and {Hayward}, Christopher C. and {Hirschmann}, Michaela and {Ilbert}, Olivier and {Kartaltepe}, Jeyhan S. and {Kocevski}, Dale D. and {Koekemoer}, Anton M. and {Kokorev}, Vasily and {Liu}, Daizhong and {Long}, Arianna S. and {McCracken}, Henry Joy and {McKinney}, Jed and {Onoue}, Masafusa and {Paquereau}, Louise and {Renzini}, Alvio and {Rhodes}, Jason and {Robertson}, Brant E. and {Shuntov}, Marko and {Silverman}, John D. and {Tanaka}, Takumi S. and {Toft}, Sune and {Trakhtenbrot}, Benny and {Valentino}, Francesco and {Zavala}, Jorge},
        title = "{COSMOS-Web: The Overabundance and Physical Nature of ``Little Red Dots''{\textemdash}Implications for Early Galaxy and SMBH Assembly}",
      journal = {The Astrophysical Journal},
         year = 2025,
        month = sep,
       volume = {991},
       number = {1},
          eid = {37},
        pages = {37},
          doi = {10.3847/1538-4357/ade984},
archivePrefix = {arXiv},
       eprint = {2406.10341},
 primaryClass = {astro-ph.GA},
       adsurl = {https://ui.adsabs.harvard.edu/abs/2025ApJ...991...37A}
}

@ARTICLE{Inayoshi_Maiolino_2025,
       author = {{Inayoshi}, Kohei and {Maiolino}, Roberto},
        title = "{Extremely Dense Gas around Little Red Dots and High-redshift Active Galactic Nuclei: A Nonstellar Origin of the Balmer Break and Absorption Features}",
      journal = {The Astrophysical Journal Letter},
         year = 2025,
        month = feb,
       volume = {980},
       number = {2},
          eid = {L27},
        pages = {L27},
          doi = {10.3847/2041-8213/adaebd},
archivePrefix = {arXiv},
       eprint = {2409.07805},
 primaryClass = {astro-ph.GA},
       adsurl = {https://ui.adsabs.harvard.edu/abs/2025ApJ...980L..27I}
}
\bibliographystyle{apsrev4-2}

\end{document}